\documentclass[aps,twocolumn,pra,superscriptaddress,amsmath,showpacs,tightenlines]{revtex4-1}
\usepackage{amssymb}
\usepackage{amsmath}
\usepackage{graphicx}
\usepackage{subfigure}
\usepackage{natbib}
\usepackage{epsfig}
\usepackage{amsfonts}
\usepackage{mathrsfs}
\usepackage{xcolor}
\usepackage[toc,page,title,titletoc,header]{appendix}

\begin{document}

\title{Interaction and entanglement engineering in driven giant atoms setup with coupled resonator waveguide}

\author{Mingzhu Weng}
\affiliation{Center for Quantum Sciences and School of Physics, Northeast Normal University, Changchun 130024, China}
\author{Xin Wang}
\affiliation{Institute of Theoretical Physics, School of Physics, Xi'an Jiaotong University, Xi'an 710049, China}
\author{Zhihai Wang}
\email{wangzh761@nenu.edu.cn}
\affiliation{Center for Quantum Sciences and School of Physics, Northeast Normal University, Changchun 130024, China}

\begin{abstract}
 We investigate the coherent interactions  mediated by the coupled resonator waveguide between two types of giant atoms. We find that the effective coupling and collective dissipation can be controlled on demand by adjusting the configuration of the giant atoms. As a result, the external driving gives birth to a substantial entanglement between two giant atoms, which exhibits a Rabi splitting character. {In the three giant atom setup, we find that the nonzero next neighbour atomic entanglement can surpass the neighbour ones, and is able to be adjust by tuning the driving phase, which serves as an artificial magnetic field.  The enhancement of next neighbour atomic entanglement can not be realized in the small atom setup.} We hope these controllable interactions in giant atom array are of great applications in the quantum information process.
\end{abstract}

\date{\today}

\maketitle

\section{introduction}
The light-matter interaction plays a crucial role in fundamental science, supporting the rapid development of quantum technology~\cite{PF2019,RG2021}. In the conventional treatment for the light-matter interactions, the atoms is usually viewed as the point-shaped dipoles~\cite{DF}. However, the demonstration of coupling between the superconducting transmon and the surface acoustic wave (SAW) has promoted the size of the matter (i.e., the transmon) to be comparable to the wavelength of the SAW~\cite{MV2014}, and the dipole approximation is broken~\cite{MV2014,AA2019,AF2021,AF2014,RM2017,AN2017,KJ2018}. Such a paradigm is subsequently named as ``giant atom'', to differ from the traditional ``small atom''. The nonlocal coupling between the giant atom and the waveguide promise to observe a lot of fancinating phenomena, such as frequency-dependent atomic relaxation rates and Lamb shifts~\cite{BK2020,YT2022,XL2022}, non-exponential atomic decay~\cite{LG2017,QY2023,LD2021}, exotic atom-photon bound
states~\cite{LG2020,XW2021,WZ2020,KH2023,WC2022}, non-Markovian decay
dynamics~\cite{GA2019,LD2023, XY2022,SG2020,XJ2023}, and chiral light-matter interactions~\cite{AF2018,AC2020,XW2022,CJ2023,XWarxiv}. Experimentally, the giant atom structures have also been realized in superconducting quantum circuits~\cite{AM2021,PY2019}, coupled waveguide arrays~\cite{SL2020} and ferromagnetic spin systems~\cite{ZQ2022}. Besides, there are some theoretical proposals in cold atoms within the optical lattices~\cite{AG2019} as well as the synthetic frequency dimensions~\cite{LD2022,HX2022}.

As for the multiple giant atoms, the waveguide can serve as a data bus, to induce the coherent interaction~\cite{XW2023arxiv}, where the geometrical configuration of the giant atom serves as a sensitive controller. For the two giant atom setup,  the braided coupling~\cite{AF2018,AM2021,AS2023}, nested coupling~\cite{AS2022,SL2021} has been predicted to be useful in some quantum information processing by constructing the decoherence-free interaction~\cite{AS2022,ERarxiv,LLarxiv} and  generating robust entanglement~\cite{BK2020,HY2021,AC2023,DD2022,QY2021,DC2020}.

In the viewpoint of the open quantum system, the wide energy band structure of the coupled resonator waveguide supplies an environment for the giant atom, to introduce the possible individual and collective atomic dissipation. One of the actionable concerned topics is the non-equilibrium dynamics of the open system which is subject to the external driven and dissipation~\cite{LM2013,TL2017,TF2018,FM2018,LR2022,NP2022}. Especially, in the giant atoms system where the dissipation can be controlled by the geometrical configuration of the giant atom~\cite{HY2021,AC2023,XY2023}.

In this paper, we tackle the above issues by considering an array of giant atoms which couple to the coupled resonator waveguide. After tracing out the degree of freedom of the waveguide, we reach three cases of effective interaction and collective dissipation  among the giant atoms. By controlling the strength and the phase of the driving fields, we find that the steady state entanglement can be engineered on demand. More interestingly, we find that there is the cyclic energy diagram in three giant atom setup. This allows the phase of the external driving generated by the artificial gauge field~\cite{PR2017,MA2013,LT2014,IB2012,YY2021,AL2012,AL2016} to enhance the next neighbour atomic entanglement so as to surpass the neighbour atomic entanglement. It is not possible in the small atoms counterpart.

The rest of the paper is organized as follows. In Sec.~\ref{model}, we describe our model and discuss the controllable effective coupling and collective dissipation induced by the waveguide. In Sec.~\ref{twoatom}, we discuss the dynamic behavior and the steady-state entanglement of the system in the two giant atom configuration. In Sec.~\ref{threeatom}, we generalize to the three giant atom setup and the steady-state entanglement behavior of the system was discussed in comparison with the small atom configuration. In Sec.~\ref{con}, we provide a short summary and discussion. Some detailed derivations of the master equation for giant atoms and small atoms are given in the Appendixes.

\section{Model and Master equation}
\label{model}
\begin{figure}
\begin{centering}
\includegraphics[width=1\columnwidth]{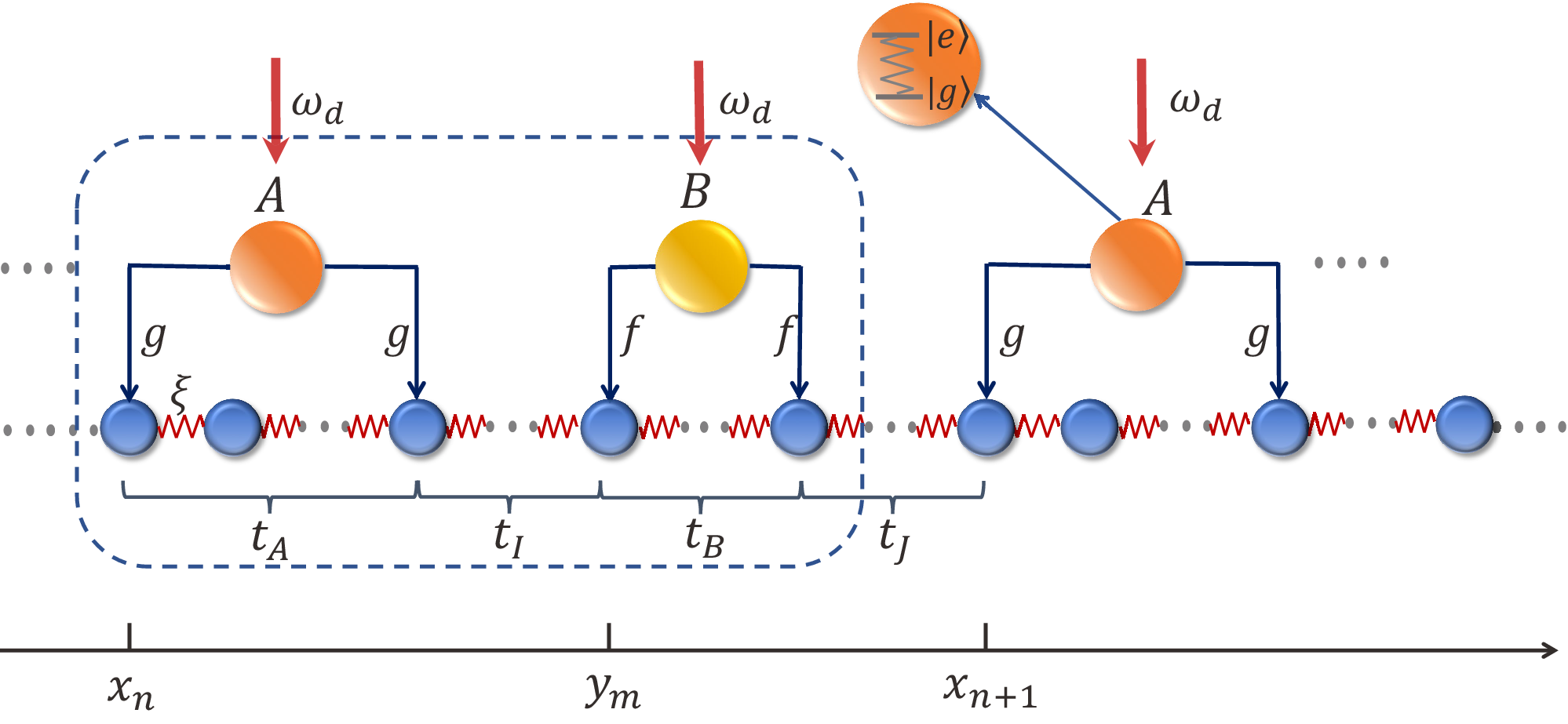}
\caption{Sketch of giant atoms coupled to a 1D coupled-resonator waveguide, where two types of giant atoms are arranged alternately. The odd (even) number of giant atoms are labeled as $A$ $(B)$ represented by orange ball (yellow ball). }
\label{device}
\end{centering}
\end{figure}

We consider an array of two-level giant atoms which interacts with a one-dimensional coupled-resonator waveguide as sketched in Fig.~\ref{device}. The giant atomic array is composed of two types of giant atoms (labeled $A$ and $B$ in Fig.~\ref{device}) which are arranged alternately with numbers being $N$ and $M$, respectively, and each giant atom couples to the waveguide at two sites. The Hamiltonian of the system is written as $H=H_c+H_a+H_I\,(\hbar=1)$,
\begin{eqnarray}
H_{c}
&=&\omega_{c}\underset{j}{\sum}a_{j}^{\dagger}a_{j}
-\xi\underset{j}{\sum}(a_{j+1}^{\dagger}a_{j}+a_{j}^{\dagger}a_{j+1}), \\H_a&=&\frac{\Omega}{2}\underset{n}{\sum}\sigma_{z}^{(n)}+\frac{\Omega}{2}\underset{m}{\sum}\Gamma_{z}^{(m)},\\
H_I
&=&g\underset{n}{\sum}[(a_{{x_{n}}}^{\dagger}+a_{{x_{n}+t_{A}}}^{\dagger})\sigma_{-}^{(n)}+{\rm H.c.}]\nonumber \\
&&+f\underset{m}{\sum}[(a_{{y_{m}}}^{\dagger}+a_{{y_{m}+t_{B}}}^{\dagger})\Gamma_{-}^{(m)}+{\rm H.c.}].
\end{eqnarray}
Here $\omega_c$ is the frequency of the resonators, $a_j$ is the annihilation operator on site $j$, $\xi$ refer to the neighbour coupling strength. The spacing between all neighbouring resonators in the coupled-resonator waveguide is the same, hereafter we take it as the unit of length. $\sigma_\pm (\Gamma_\pm)$ are the Pauli operators of the giant atom $A (B)$, $\Omega$ is the transition frequency of the giant atom between the ground state $|g\rangle$ and the excited state $|e\rangle$. Note that $g,f$ are the coupling strength of $A$ and $B$ giant atoms with the waveguide, respectively. $t_A(t_B)$ characterizing the size of the giant atom $A(B)$.  Giant atom $A$ in the $n$th cell is coupled to the waveguide at position $x_n$ and $x_n+t_A$, $n=1,2,\ldots,N$. Similarly, the coupling position of $B$ giant atom within the same cell are $y_m$ and $y_m+t_B$, $m=1,2,\ldots,M$. $y_{i}=x_{i}+t_{A}+t_{I}$, $x_{i}=y_{i-1}+t_{B}+t_{J}$ with the parameter $t_I$, $t_J$ being the intra-cell atomic distance and the extra-cell atomic distance, respectively.

Considering the length of the waveguide $N_c\rightarrow\infty$, we can rewrite the Hamiltonian of the system in momentum space. By introducing the Fourier transformation $a_j=\sum_{k}a_{k}e^{ikj}/\sqrt{N_c}$, the Hamiltonian $H_c$ becomes $H_k=\sum_{k}\omega_ka^{\dagger}_{k}a_{k}$ with a dispersion relation is given by $\omega_k=\omega_c-2\xi\cos{k}$. Therefore, the waveguide supports a single-photon continual band with center frequency $\omega_c$ and a bandwidth $4\xi$. In the momentum space, the atom-waveguide coupling Hamiltonian $H_I$ is expressed as
\begin{eqnarray}
H_{I}&=&	g\frac{1}{\sqrt{N_{c}}}\underset{n,k}{\sum}[(a_{k}^{\dagger}e^{-ikx_{n}}+a_{k}^{\dagger}e^{-ik(x_{n}+t_{A})})\sigma_{-}^{(n)}+{\rm H.c.}]\nonumber \\	&&+f\frac{1}{\sqrt{N_{c}}}\underset{m,k}{\sum}[(a_{k}^{\dagger}e^{-iky_{m}}+a_{k}^{\dagger}e^{-ik(y_{m}+t_{B})})\Gamma_{-}^{(m)}+{\rm H.c.}].\nonumber\\
\
\end{eqnarray}

\begin{figure*}
\begin{centering}
\includegraphics[width=2.05\columnwidth]{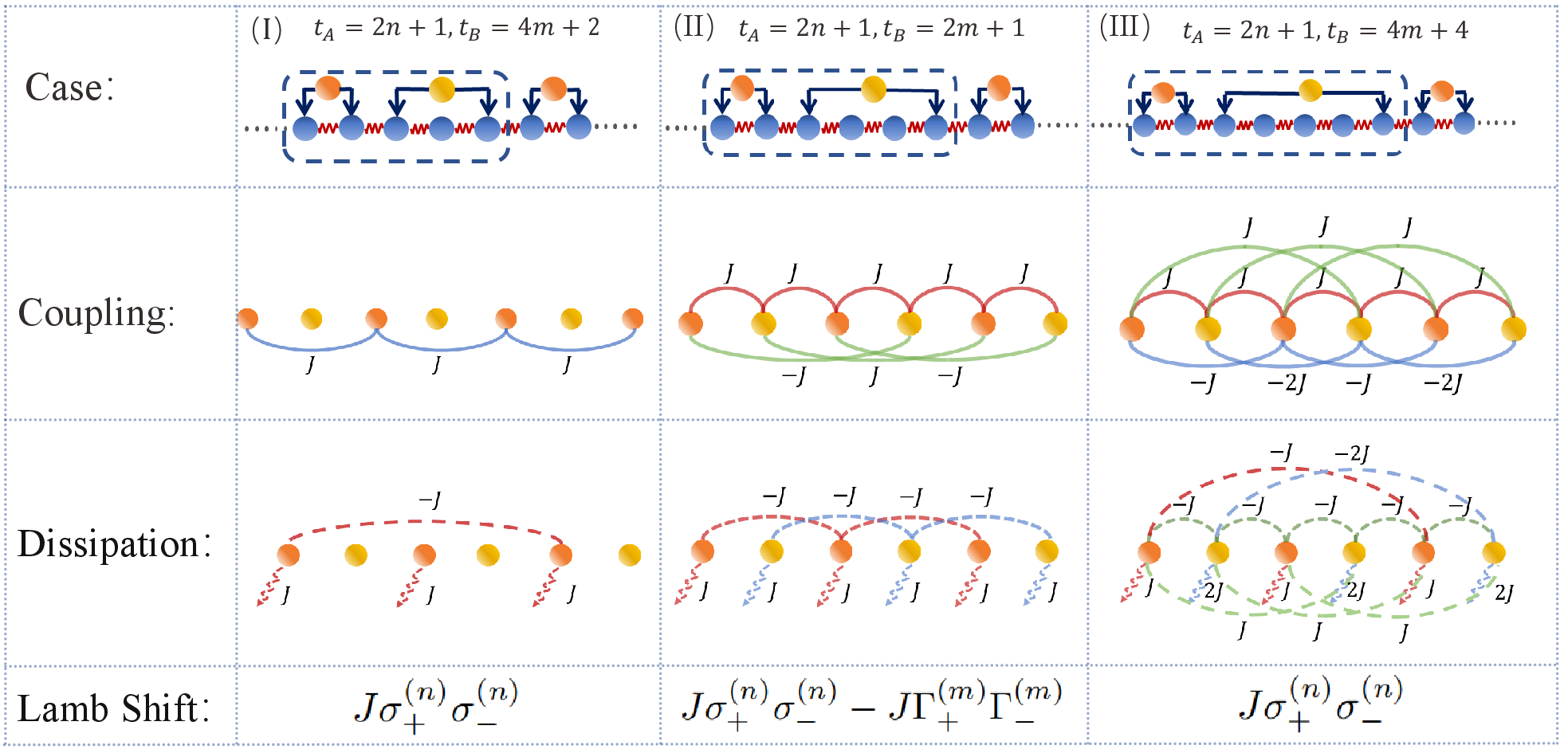}
\caption{The effective couplings and dissipations for three different cases. The first line is the schematic illustration of different systems of the odd (even) number of atoms are labeled as  $A$ $(B)$ and represented by orange (yellow) ball. The second line corresponds to a schematic diagram of the corresponding effective coherent interaction, that is, the neighbour interaction (red solid line), the next neighbour interaction (blue solid line) and secondary neighbour interaction (green solid line). The third line shows the diagram of collective dissipation, in which the dissipation between $A$ atoms is represented by the red dashed line, the dissipation between $B$ atoms is represented by the blue dashed line, and the collective dissipation between $A-B$ atoms is represented by the green dashed line.
{The red wave line is the independent dissipation induced by the waveguide.}
{The last line is the Lamb shift induced the waveguide. $J=g^{2}/\xi $ with $g=f$ being the coupling strength of giant atoms and waveguide.}}
\label{masterFig}
\end{centering}
\end{figure*}

{Let us first consider the weak-coupling or broadband limit $g,f\ll\xi$.} In this regime, the waveguide modes can be eliminated by adopting the Born-Markov approximation. To obtain the master equation for the density matrix of the giant atoms, we work in the momentum representation and  the interaction picture. Then, the interaction Hamiltonian $H_{I}$ becomes\cite{GC2016}
\begin{eqnarray}
H_{I}&=&g\underset{n}{\sum}[\sigma_{+}^{(n)}E(x_{n},t)e^{i\Omega t}+\sigma_{+}^{(n)}E(x_{n}+t_{A},t)e^{i\Omega t}+{\rm H.c.}]\nonumber\\
&&+f\underset{m}{\sum}[\Gamma_{+}^{(m)}E(y_{m},t)e^{i\Omega t}+\Gamma_{+}^{(m)}E(y_{m}+t_{B},t)e^{i\Omega t}\nonumber
\\
&&+{\rm H.c.}].
\end{eqnarray}
where $E(X,t)=\frac{1}{\sqrt{N_{c}}}\underset{k}{\sum}e^{-i\omega_{k}t}e^{ikX}a_{k}$ is the field operator at site $X$ and the master equation is formally written as~\cite{HB}
\begin{equation}
\frac{d}{dt}{\rho}(t)=-\int_{0}^{\infty}d\tau{\rm Tr_{c}}\{[H_{I}(t),[H_{I}(t-\tau),\rho_{c}\otimes\rho(t)]]\}.
\end{equation}

In the following, we will consider that the giant atoms are resonant with the bare resonator, that is, $\Omega=\omega_c$. As a result, we obtain a master equation for the reduced density operator of the giant atoms (the  detailed calculations are given in Appendix.~\ref{A1}) as
\begin{eqnarray}
\overset{\cdot}{\rho}&=&-i[\mathcal{H},\rho]\nonumber\\
&&+\underset{n,m}{\sum}[g^{2}U_{11}^{(n,m)}(2\sigma_{-}^{(n)}\rho\sigma_{+}^{(m)}-\sigma_{+}^{(m)}\sigma_{-}^{(n)}\rho-\rho\sigma_{+}^{(n)}\sigma_{-}^{(m)})\nonumber\\
&&+f^{2}U_{22}^{(n,m)}(2\Gamma_{-}^{(n)}\rho\Gamma_{+}^{(m)}-\Gamma_{+}^{(n)}\Gamma_{-}^{(m)}\rho-\rho\Gamma_{+}^{(n)}\Gamma_{-}^{(m)})\nonumber\\
&&+gfU_{12}^{(n,m)}(2\sigma_{-}^{(n)}\rho\Gamma_{+}^{(m)}-\sigma_{+}^{(n)}\Gamma_{-}^{(m)}\rho-\rho\sigma_{+}^{(n)}\Gamma_{-}^{(m)})\nonumber\\
&&+gfU_{21}^{(n,m)}(2\Gamma_{-}^{(n)}\rho\sigma_{+}^{(m)}-\Gamma_{+}^{(n)}\sigma_{-}^{(m)}\rho-\rho\Gamma_{+}^{(n)}\sigma_{-}^{(m)})].\nonumber\\
\
\label{masterequation}
\end{eqnarray}

where the coherent coupling between the atoms are described by the Hamiltonian
\begin{eqnarray}
\mathcal{H}&=&\underset{n,m}{\sum}(\frac{\Omega}{2}\sigma_{z}^{(n)}+\frac{\Omega}{2}\Gamma_{z}^{(m)})\nonumber\\ &&+\underset{n,m}{\sum}[g^{2}I_{11}\sigma_{+}^{(n)}\sigma_{-}^{(m)}+f^{2}I_{22}\Gamma_{+}^{(n)}\Gamma_{-}^{(m)}]\nonumber\\ &&+\underset{n,m}{\sum}[gf(I_{12}\sigma_{+}^{(n)}\Gamma_{-}^{(m)}+I_{21}\Gamma_{+}^{(n)}\sigma_{-}^{(m)})].
\label{effectiveH}
\end{eqnarray}
In the above equations, we have defined $U_{ij}=\mathrm{Re}(A_{ij}),I_{ij}=\mathrm{Im}(A_{ij})(i,j=1,2)$ with
\begin{eqnarray}
A_{11}&=&\underset{n}{\sum}\underset{m}{\sum}\frac{1}{2\xi}(2e^{i\frac{\pi}{2}|x_n-x_m|}\nonumber\\
&&+e^{i\frac{\pi}{2}|x_n-x_m-t_A|}+e^{i\frac{\pi}{2}|x_n+t_A-x_m|}),\\
A_{22}&=&\underset{n}{\sum}\underset{m}{\sum}\frac{1}{2\xi}(2e^{i\frac{\pi}{2}|y_n-y_m|}\nonumber\\
&&+e^{i\frac{\pi}{2}|y_n-y_m-t_B|}+e^{i\frac{\pi}{2}|y_n+t_B-y_m|}),\\
A_{12}&=&\underset{n}{\sum}\underset{m}{\sum}\frac{1}{2\xi}(e^{i\frac{\pi}{2}|x_n-y_m|}+e^{i\frac{\pi}{2}|x_n-y_m-t_B|}\nonumber\\
&&+e^{i\frac{\pi}{2}|x_n+t_A-y_m|}+e^{i\frac{\pi}{2}|x_n+t_A-y_m-t_B|}),\\
A_{21}&=&\underset{n}{\sum}\underset{m}{\sum}\frac{1}{2\xi}(e^{i\frac{\pi}{2}|y_n-x_m|}+e^{i\frac{\pi}{2}|y_n-y_m-t_A|}\nonumber\\
&&+e^{i\frac{\pi}{2}|y_n+t_B-x_m|}+e^{i\frac{\pi}{2}|y_n+t_B-x_m-t_A|}).
\label{Aij}
\end{eqnarray}
From the above formula, it can be seen that the coherent interaction and the collective dissipation between the giant atoms can be modulated by the size of each giant atom, that is,  the distance between the atom-waveguide coupling points. For simplicity, we consider that the two giant atoms couple to the waveguide via same coupling strength at each point and fix the atomic spacing to be uniform by setting $t_{I}=t_{J}=1$.

\begin{figure*}
\begin{centering}
\includegraphics[width=2.05\columnwidth]{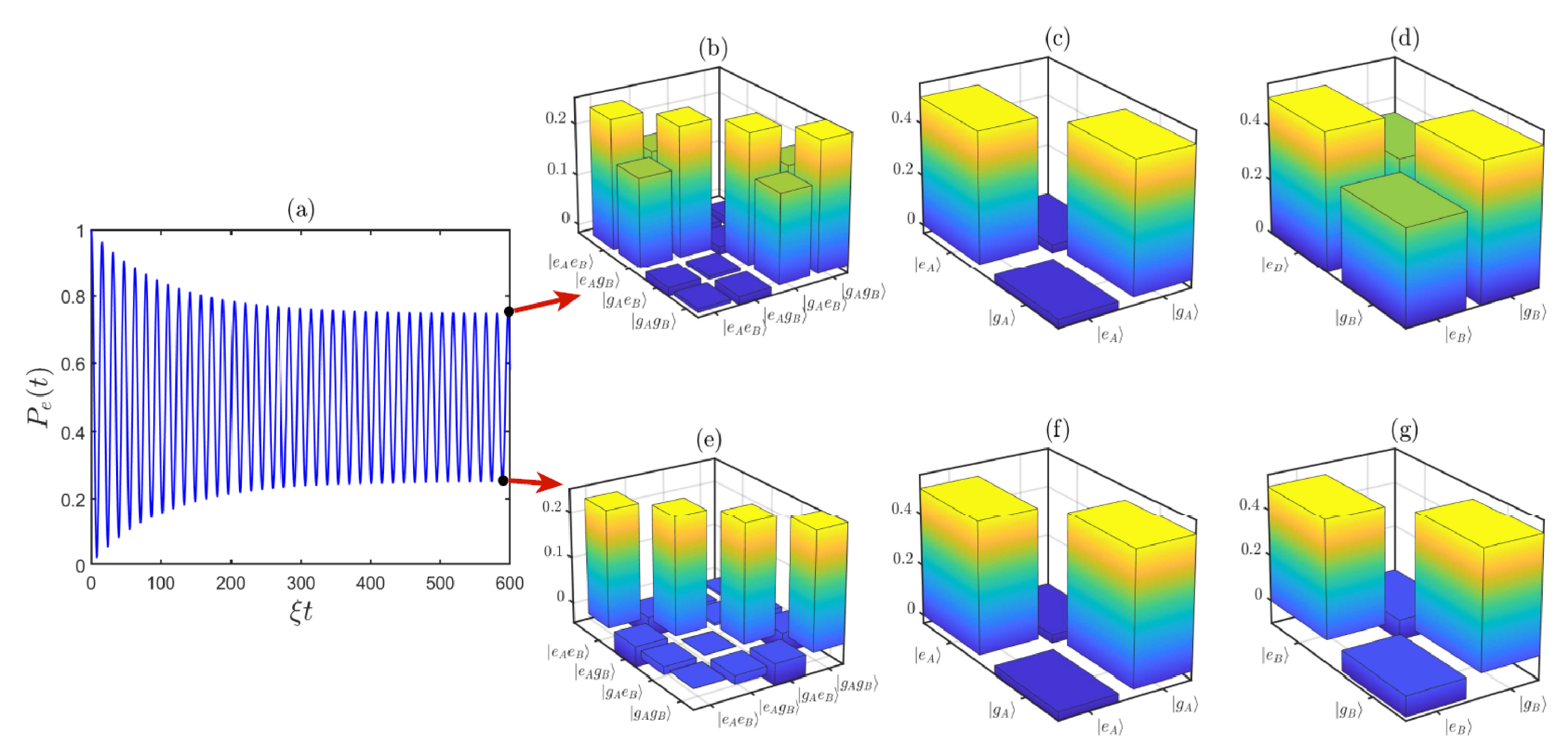}
\caption{(a) Time evolution of the average value of the Pauli operators $P_e(t)= (\langle \sigma_z \rangle+\langle \Gamma_z \rangle)/2$. (b-g) Tomography of the state of the system at different moments. The parameters are set as $\Delta=0$, $g=f=0.08\xi$ and $\eta=0.2\xi$. }
\label{twoatomfig1}
\end{centering}
\end{figure*}

By adjusting the size of the giant atoms on demand, we can obtain the following three cases of effective coherent couplings and collective dissipations, which are listed in Fig.~\ref{masterFig}.  Now, we illustrate them in detail.  Case (I): When $t_{A}=2n+1, t_{B}=4m+2$ with integral $n, m=0,1,2\ldots$, we find that $A_{22}=0,A_{12}=0,A_{21}=0$.
Therefore, the $B$ giant atoms are totally decoupled from the waveguide due to the interference effect between the two connecting points.
As for the $A$ type giant atoms, they undergo the coherent coupling, both individual and collective dissipations. Case (II): When $t_{A}=2n+1, t_{B}=2m+1$, the effective coherent couplings only exist between the different types of the atoms. Meanwhile, the collective dissipation occurs between the same types of the atoms  except for their individual dissipations. Case (III): When $t_{A}=2n+1, t_{B}=4m+4$,  both of the two types of the giant atoms undergo the individual dissipations. Moreover, the  coupling and collective dissipation exists between both of the same and different types of giant atoms. Besides the mentioned statements above, one of the most intriguing differences between the coherent coupling is that in case (III), only three atoms are needed to form the cyclic coupling, whereas in case (II), at least four are needed.

{Note that there are still three other configurations which are not mentioned in Fig.~\ref{masterFig}. They are $t_A=4n+2,t_B=4m+4$, $t_A=4n+2,t_B=4m+2$ and $t_A=4n+4,t_B=4m+4$, respectively. For $t_A=4n+2,t_B=4m+4$ and $t_A=4n+2,t_B=4m+2$, both of the two types of the giant atoms are totally decoupled from the waveguide with $A_{11}=A_{22}=A_{12}=A_{21}=0$. When $t_A=4n+4,t_B=4m+4$, it has the same form of the coherence coupling form as case (II) in Fig.~\ref{masterFig}. Therefore, we will not discuss them in what follows.
Based on the three cases listed in Fig.~\ref{masterFig}, we discuss the dynamical behavior for two and three giant atoms setup.}
The results for the small atom setup will be listed in Appendix.~\ref{A2} for comparison.

\section{Two Giant Atoms}
\label{twoatom}

In this section, we first discuss the setup with only two giant atoms setup, which are labeled by $A$ and $B$, respectively. A general master equation
can be written as
\begin{eqnarray}
\dot{\rho}&=&-i[\mathcal{H}+H_{d},\rho]\nonumber \\
&&+J_1D_{[\sigma_{+},\sigma_{-}]\rho}+J_2D_{[\Gamma_{+},\Gamma_{-}]\rho}\nonumber \\
&&+J_3\left(D_{[\sigma_{+},\Gamma_{-}]\rho}+D_{[\Gamma_{+},\sigma_{-}]\rho}\right),
\end{eqnarray}
where $\mathcal{H}$ is the effective Hamiltonian, and $D_{[O_1,O_2]\rho}=2O_2\rho O_1 -\rho O_1O_2-O_1O_2 \rho$. The Hamiltonian $\mathcal{H}$ and the coefficients $J_i\,(i=1,2,3)$ depend on the configuration of the giant atom. In the rotating frame, the driving Hamiltonian is written as $H_d=\eta(\sigma_{+}+\Gamma_{+}+{\rm H.c.})$.
{Here, the driving field is applied directly to the atoms, and does not cause the interactions between the atoms.}
$\mathcal{H}_{\rm I}(\mathcal{H}_{\rm II},\mathcal{H}_{\rm III})$ corresponds to the effective Hamiltonian for case (I) (case (II), case (III)).

For the (I) case of $t_{A}=2n+1, t_{B}=4m+2$ with integral $n, m=0,1,2\ldots$, the Hamiltonian is
\begin{equation}
\mathcal{H}_{\rm {I}}=\Delta \sigma_{z}+\Delta \Gamma_{z}+J\sigma_{+}\sigma_{-}
\end{equation}
where $J=g^2/\xi$ and $\Delta=\Omega-\omega_d$ is the detuning between the frequency of the giant atom and the frequency of driving field.  The coefficients in the dissipation terms satisfy {$J_1=J$, and $J_i=0$ for $i=2,3$}. We now consider that the two giant atoms are both excited initially ($|\psi(0)\rangle=|e,e\rangle$) and explore the time evolution of the atomic populations. In Fig.~\ref{twoatomfig1}(a), We find that the system cannot reach a steady state even when the evolution time tends to be infinity.
The distinguished dynamics for the giant atoms can be explained by the cartoon coupling scheme in the first column of Fig.~\ref{masterFig}. The two giant atoms are isolated from each other with neither effective coupling nor collective decay. The giant atom $A$ experiences both of the dissipation induced by the waveguide and the external driving field, and eventually achieves its steady state. On the contrary, the giant atom $B$ is subject only to the external driving field, and exhibits the Rabi oscillation.
We further pick two points $\xi t=597$ and $\xi t=589$ among the moments during the time evolution when the atomic population achieves its maximum and minimum values in Fig~\ref{twoatomfig1}(a), and tomograph the corresponding quantum state in Figs.~\ref{twoatomfig1}(b) and (e).
It shows that, the population exhibits an uniform distribution diagonally for both of two states.
{However, unlike the obvious coherence in the bare basis of the two atom $\rho_{e_Ag_B,e_Ae_B}=0.1712$ as shown in Fig.~\ref{twoatomfig1}(b), the coherence of the steady state is $\rho_{e_Ag_B,e_Ae_B}=0.04$ shown in Fig.~\ref{twoatomfig1}(e). The later one is only $0.2$ times of the former one, so the coherence shown in Fig.~\ref{twoatomfig1}(e)  can be neglected.}
In Figs.~\ref{twoatomfig1}(c,d,f,g), we further illustrate the tomography of the reduced density matrix $\rho_{A(B)}={\rm Tr}_{B(A)}\rho$ in the two states in order to investigate the states of the atoms $A$ and $B$, respectively.
At these two moments, the state of giant atom $A$ is same with each other as shown in Figs.~\ref{twoatomfig1}(c) and (f), but the giant atom $B$ behaves differently as shown in Figs.~\ref{twoatomfig1}(d) and (g) in which the coherence shows the similar results with that in Figs.~\ref{twoatomfig1}(a) and (e), respectively.
{Therefore, the unstable behavior originates from the isolated giant atom $B$, which is continuously driven by the external driving field.}

\begin{figure}
\includegraphics[width=1\columnwidth]{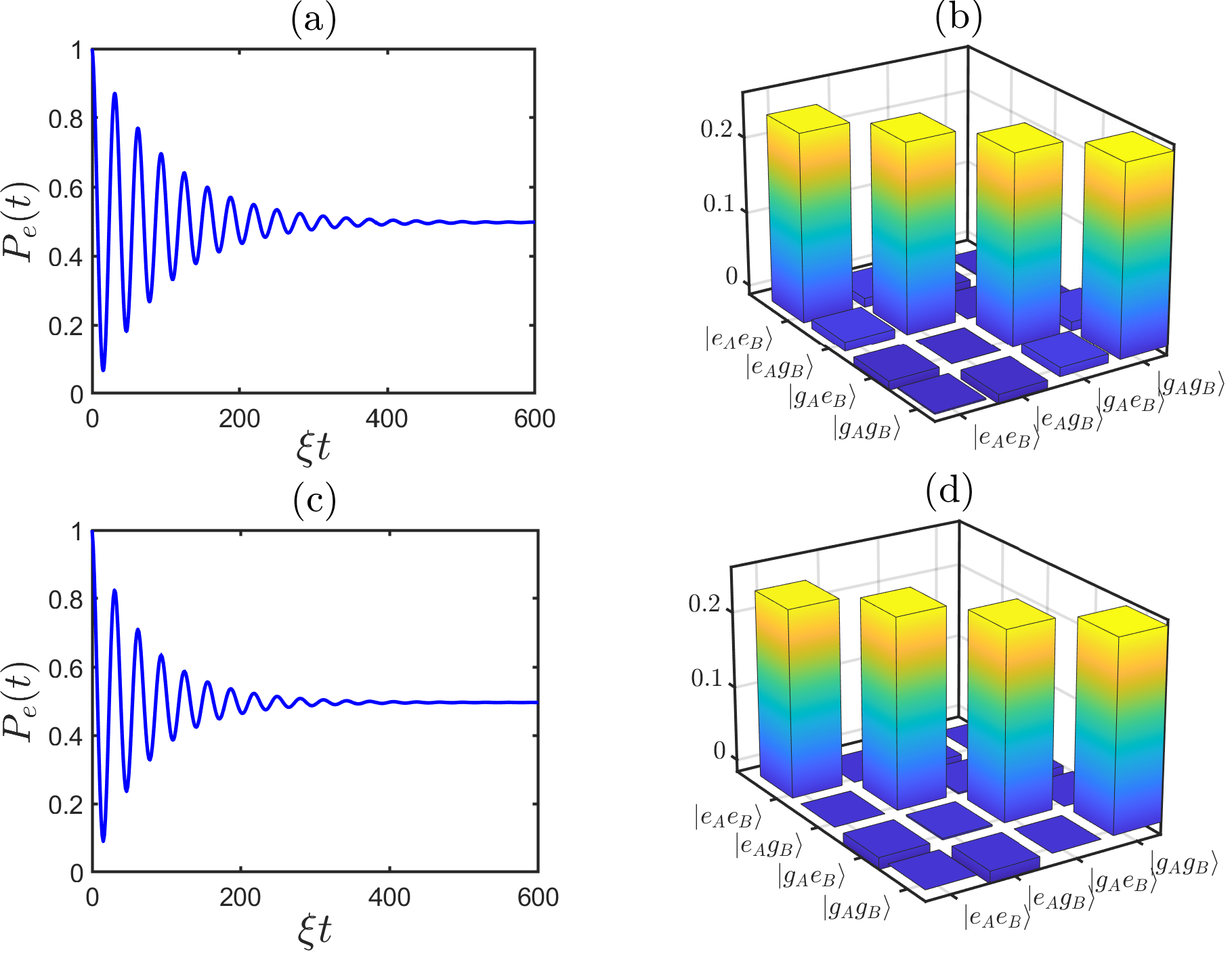}
\caption{ The giant atomic population evolution (a,c) and the tomography of the steady state (b,d).  The results for cases (II) and (III) are demonstrated in (a,b) and (c,d) respectively. The parameters are set as $\Delta=0$, $g=f=0.08\xi$ and $\eta=0.2\xi$.}
\label{twoatomfig2}
\end{figure}

As shown in Fig.~\ref{masterFig}, the effective Hamiltonians for (II) and (III) cases become
\begin{eqnarray}
\mathcal{H}_{\rm {II}}&=&\Delta\sigma_{z}+\Delta\Gamma_{z}+J\sigma_{+}\sigma_{-}
-J\Gamma_{+}\Gamma_{-}\nonumber\\
&&+J(\sigma_{+}\Gamma_{-}+\Gamma_{+}\sigma_{-}).
\end{eqnarray}
\begin{equation}
\mathcal{H}_{\rm {III}}=\Delta\sigma_{z}+\Delta\Gamma_{z}+J\sigma_{+}\sigma_{-}+
J(\sigma_{+}\Gamma_{-}+\Gamma_{+}\sigma_{-}).
\end{equation}
The effective coupling is the identical in both cases, only the Lamb shift term induced by the waveguide is different. As for the dissipation counterpart, we have $J_1=J_2=J,J_3=0$ for case (II) and $J_1=J,\, J_2=2J,\, J_3=-J$ for case (III), respectively. Due to the similarity between the above two effective Hamiltonian $\mathcal{H}_{\rm II}$ and $\mathcal{H}_{\rm III}$, we observe that the system will undergo an initial oscillation and finally reach the steady state in (II) and (III) cases, whose dynamics are demonstrated in Figs.~\ref{twoatomfig2} (a) and (c), respectively.
For the steady state, the tomography results in Figs.~\ref{twoatomfig2} (b) and (d) shows that they are almost the maximum mixed state without obvious coherence terms.

\begin{figure}
\includegraphics[width=1\columnwidth]{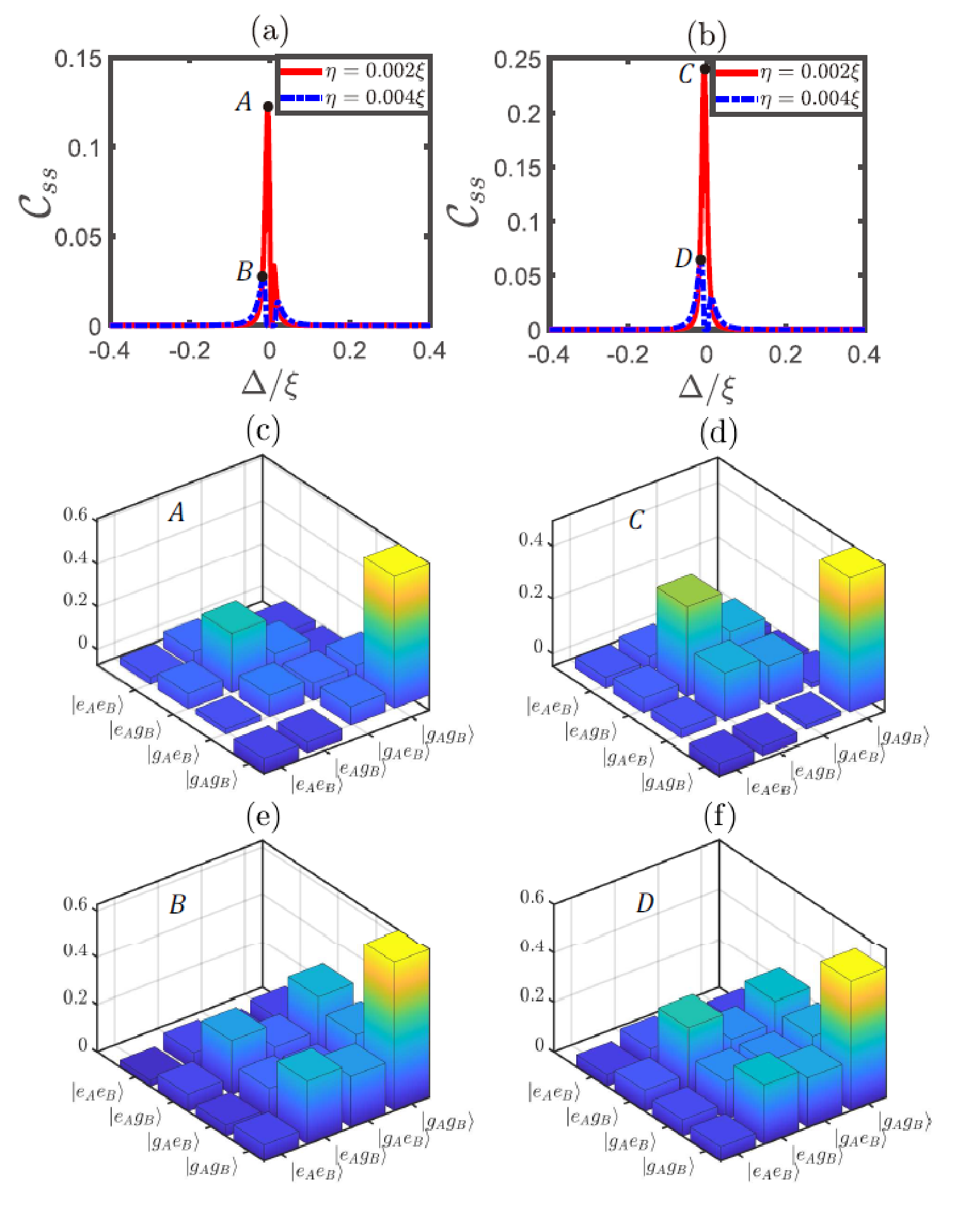}
\caption{The concurrence of the steady states for the case (II) in (a) and the case (III) (b). The parameters are set as $\Delta=0$ and $g=f=0.05\xi$. (c-f) are the tomography of the state at the four black dots in (a) and (b).}
\label{twoatomfig3}
\end{figure}

{We furthermore explore the steady-state entanglement for case (II) and case (III), which are able to reach the steady state. The steady-state entanglement is quantified by the concurrence of the state proposed by Hill and Wootters~\cite{SA1997}.
For stronger driving strengths, as can be obtained from the tomography in Fig.~\ref{twoatomfig2}(b,d), the system is weakly coherent and being in the maximum mixing state but without entanglement.
Therefore, we explore the steady-state entanglement when the driving strength is in the same order as the effective coupling strength $J$.}
The results for (II) and (III) cases are demonstrated in
Figs.~\ref{twoatomfig3} (a) and (b), respectively under different driving strength.
When the driving strength is $\eta=0.004\xi$, we find the appearance of the Rabi splitting in both of the cases, which indicates the waveguide induced effective coupling between the two giant atoms.
{When the driving field is resonant with the frequency difference between the dressed state and the ground state, a relative large entanglement can be achieved and the concurrence peaks locate at $\delta=\pm \sqrt{2}J$.}
However, when the system is subject to a weakly driven of $\eta=0.002\xi$, there are two peaks in case (II) while they are fused into a single one in case (III). This is because the  dissipation rate of the giant atom $B$ is $J_{2}=2J$, which is larger than that in case (II) with $J_{2}=J$.
{On the other hand, due to the waveguide induced dissipation, the steady state is a mixed state of the dressed state and the ground state, therefore the concurrence is much smaller than $1$ as shown in the Figs.~\ref{twoatomfig3} (a) and (b).
To demonstrate the state more clearly, we add the tomography of steady states in Figs.~\ref{twoatomfig3}(c-f).  When $\eta=0.002\xi$, as shown in Figs.~\ref{twoatomfig3}(c,d), the probability of being in the dressed state is greater in Fig.~\ref{twoatomfig3}(d) than that in Fig.~\ref{twoatomfig3}(c).
Correspondingly, the entanglement in Fig.~\ref{twoatomfig3}(b) is larger than that in Fig.~\ref{twoatomfig3}(a).
Similarly, for $\eta=0.004\xi$, the greater coherence in case (III) also corresponds to a stronger entanglement.}

\section{Three Giant Atoms}
\label{threeatom}
\begin{figure}
\includegraphics[width=1.01\columnwidth]{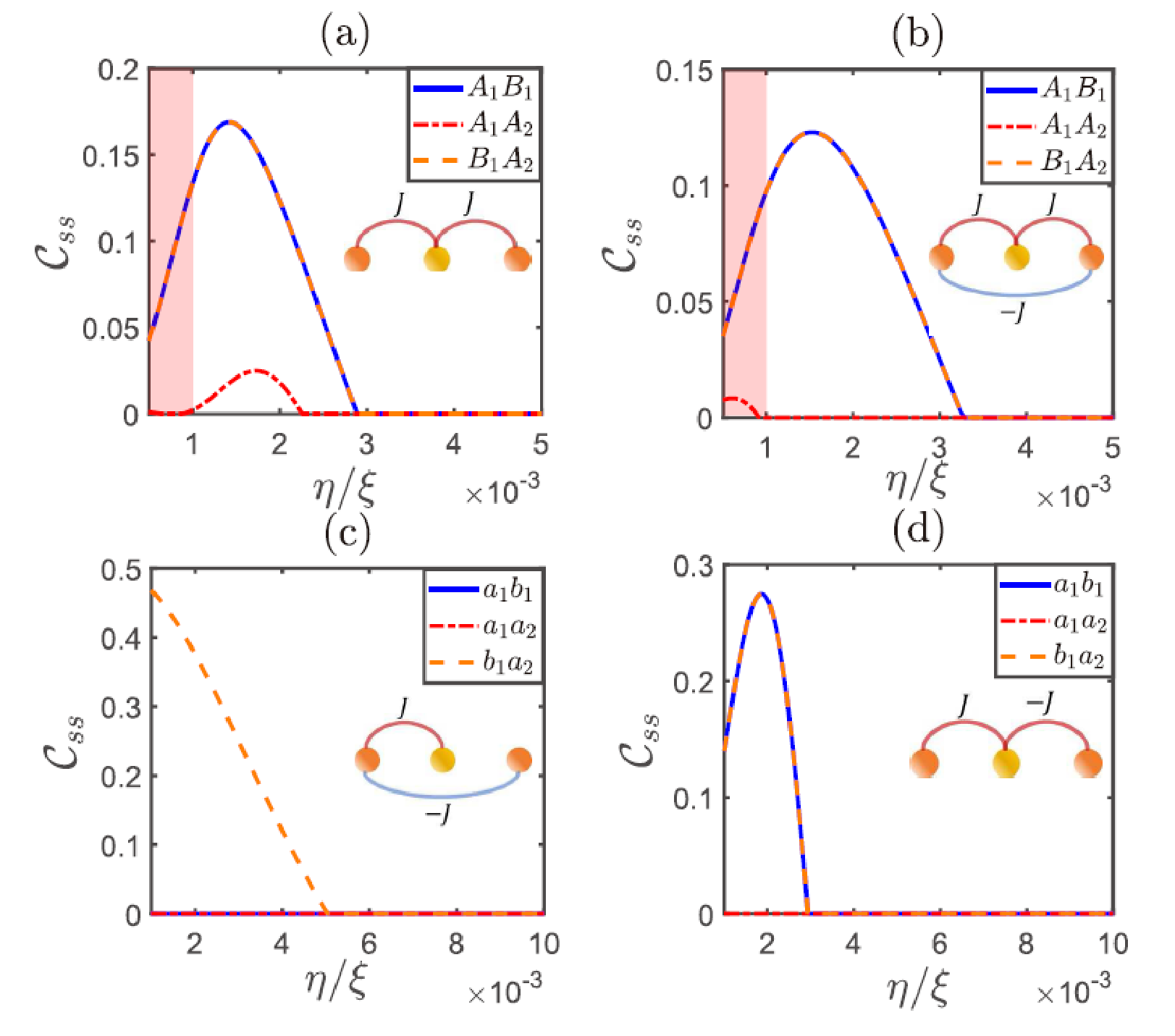}
\caption{(a) and (b): The steady-state entanglement in the three giant atom setting with the case (II) and case (III), respectively. (c) and (d): The steady-state entanglement of the small atom setup. The red dashed line gives the results for the {next neighbour} entanglement as a function of drive strength. The parameters are set as $\Delta=0$, $g=f=0.05\xi$.}
\label{threeatomfig}
\end{figure}

To explore the potential application of our proposal with giant atoms in the quantum information processing, we generalize the above discussion to the setup consisting of three giant atoms which are composed of two $A$ type atoms (denoted by $A_1$ and $A_2$, respectively) and one $B$ type atom (denoted by $B_1$).
Since the $B$ type atom in case (I) is decoupled with the waveguide, we here constraint ourselves in case (II) and case (III).
Then, the master equation for the giant atoms are
$\dot{\rho}=-i[\mathcal{H}_i,\rho]+D_i\rho\,(i={\rm II}, {\rm III})$, where the corresponding effective Hamiltonians and the dissipative terms in the rotating frame are respectively
\begin{eqnarray}
\mathcal{H}_{\rm II}&=&\Delta(\sigma_{z}^{(1)}+\sigma_{z}^{(2)})
+\Delta\Gamma_{z}^{(1)}+J(\sigma_{+}^{(1)}\sigma_{-}^{(1)}
+\sigma_{+}^{(2)}\sigma_{-}^{(2)})\nonumber\\
&&-J\Gamma_{+}^{(1)}\Gamma_{-}^{(1)}+J(\sigma_{+}^{(1)}
\Gamma_{-}^{(1)}+\Gamma_{+}^{(1)}\sigma_{-}^{(2)}+{\rm H.c.}),\\
\label{HThreeII}
D_{\rm II}\rho&=&JD_{[\sigma_{+}^{(1)},\sigma_{-}^{(1)}]\rho}+JD_{[\sigma_{+}^{(2)},\sigma_{-}^{(2)}]\rho}+JD_{[\Gamma_{+}^{(1)},\Gamma_{-}^{(1)}]\rho}\nonumber\\
&&-J\left(D_{[\sigma_{+}^{(1)},\sigma_{-}^{(2)}]\rho}+D_{[\sigma_{+}^{(2)},\sigma_{-}^{(1)}]\rho}\right),
\end{eqnarray}

and
\begin{eqnarray}
\mathcal{H}_{\rm III}&=&\Delta(\sigma_{z}^{(1)}+\sigma_{z}^{(2)})+\Delta\Gamma_{z}^{(1)}+
J(\sigma_{+}^{(1)}\sigma_{-}^{(1)}+\sigma_{+}^{(2)}\sigma_{-}^{(2)})\nonumber\\
&&+J(\sigma_{+}^{(1)}\Gamma_{-}^{(1)}+\Gamma_{+}^{(1)}\sigma_{-}^{(2)}-\sigma_{+}^{(1)}\sigma_{-}^{(2)}+{\rm H.c.}),\\
D_{\rm III}\rho&=&JD_{[\sigma_{+}^{(1)},\sigma_{-}^{(1)}]\rho}+JD_{[\sigma_{+}^{(2)},\sigma_{-}^{(2)}]\rho}+2JD_{[\Gamma_{+}^{(1)},\Gamma_{-}^{(1)}]\rho}\nonumber\\
&&-J\left(D_{[\sigma_{+}^{(1)},\Gamma_{-}^{(1)}]\rho}+D_{[\Gamma_{+}^{(1)},\sigma_{-}^{(1)}]\rho}\right)\nonumber\\
&&-J\left(D_{[\sigma_{+}^{(2)},\Gamma_{-}^{(1)}]\rho}+D_{[\Gamma_{+}^{(1)},\sigma_{-}^{(2)}]\rho}\right).
\end{eqnarray}

In Figs.~\ref{threeatomfig}(a) and (b), we plot the two atom steady state entanglement as a function of the {drive strength} for case (II) and (III), respectively. Correspondingly, the waveguide induced effective atomic couplings are demonstrated inset. In both cases, the entanglement between the adjacent atom ($A_1B_1$ and $B_1A_2$) agree with each other, and is larger than that of the next neighbour ones. The difference is reflected in the next neighbour atomic entanglement.
{For weak driving, a more pronounced next neighbour entanglement can be generated in case (III) as shown in the shaded regime of Figs.~\ref{threeatomfig}(a,b).}
The significant difference stem from the waveguide induced atomic coupling. In case (II), there is only the neighbour atomic coupling with strength $J$.
For case (III), we further find the next neighbour atomic coupling with strength $-J$ and it forms a cyclic coupling.
Therefore, compare to case (II), there is more next-neighbour atomic entanglement in case (III).
{When the driving strength is further increased to be greater than the effective coupling strength induced by the waveguide, the neighbour entanglement vanishes that the system is in a maximally mixed state, which is consistent with the discussion in Figs.~\ref{twoatomfig2} and~\ref{twoatomfig3}.}

As a comparsion, we also discuss the steady state entanglement when an array of small atoms couple to the waveguide with one coupling point for each atom. Here, we consider  that two neighboring small atoms (labeled $a$ and $b$)form a unit cell and both of the intra-cell and extra-cell atomic distances can be adjusted on demand.
The Hamiltonian of the small atom setup and the master equation is given in Appendix.~\ref{A2}. There exist two coupling cases which are listed in Fig.~\ref{smallinteraction}. Correspondingly, the steady state concurrence as a function of the driving strength  and the effective inter atom coupling are plotted in Figs.~\ref{threeatomfig}(c) and (d). It shows that, for both of the two cases, the next-neighbour atomic entanglement is always zero, due to the absence of the cyclic coupling configuration.

\section{Entanglement manipulation by artificial magnetic field}

\begin{figure}
\includegraphics[width=1\columnwidth]{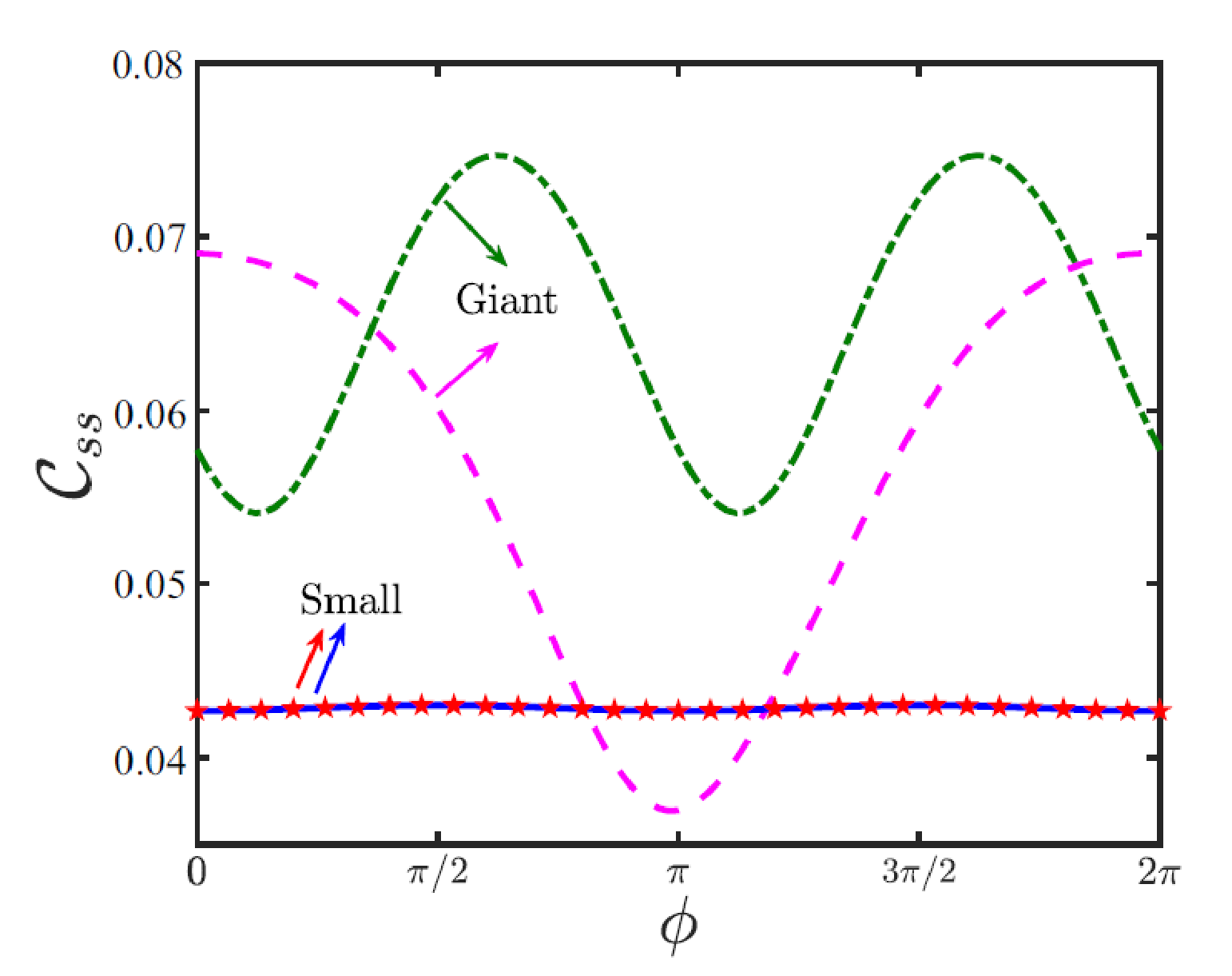}
\caption{Steady-state entanglement modulated by artificial magnetic field for the case of the giant atoms and the small atoms. The green dashed line and purple dashed line correspond to the case (II) and case (III) geometric configurations in the giant atom configuration, respectively. The blue solid line and red star line represent the results for small atoms.  The parameters are set as $\Delta=0$, $g=f=0.06\xi$ and $\eta=0.002\xi$.}
\label{magtwo}
\end{figure}

\begin{figure*}
\begin{centering}
\includegraphics[width=2.06\columnwidth]{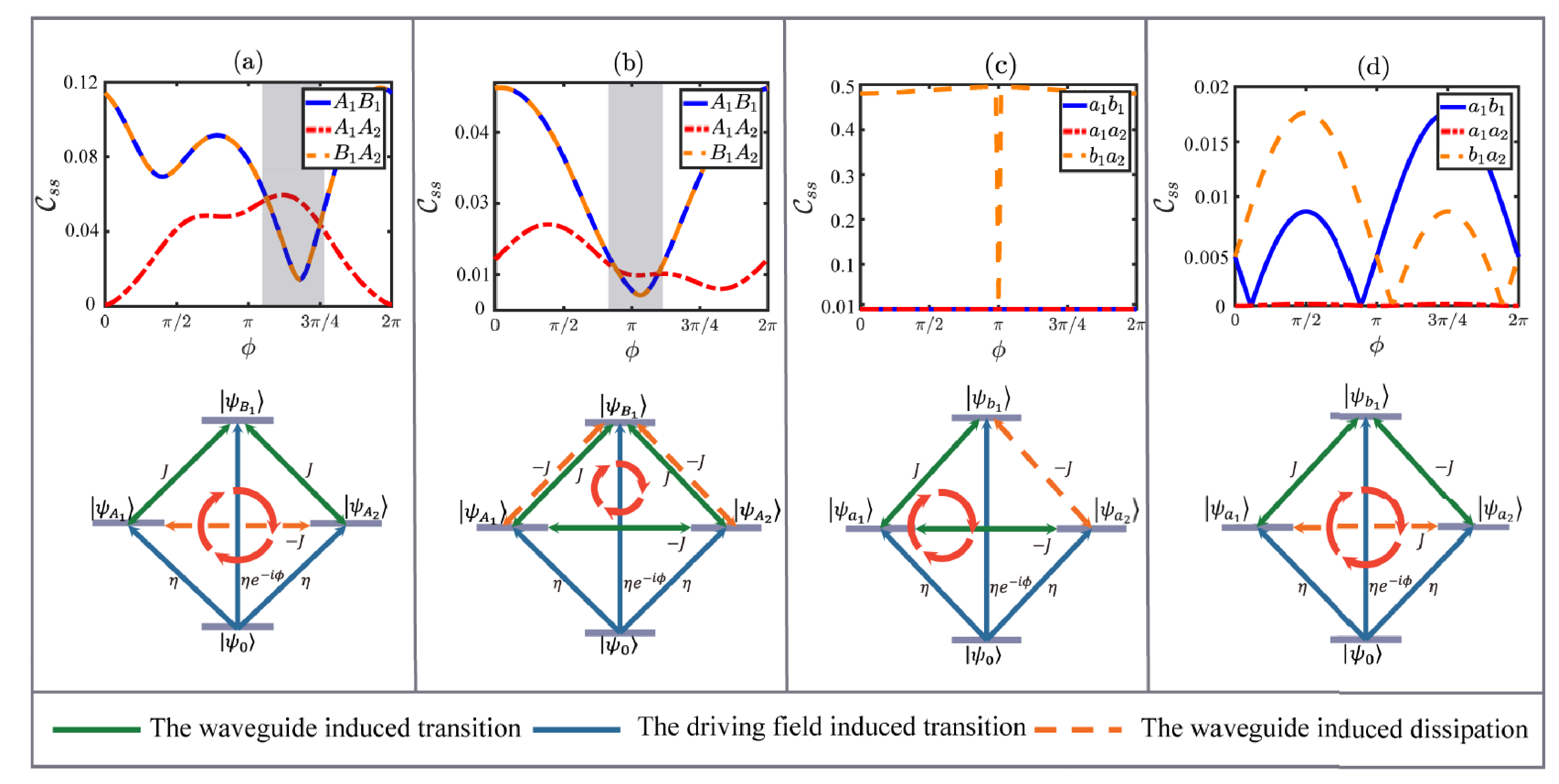}
\caption{The steady-state entanglement modulated by artificial magnetic field for the case of the giant atoms (a-b) and the small atoms (c-d). The red dashed line represent the next neighbour entanglement. The blue solid line and orange dashed line represent the neighbour entanglement. The parameters are set as $\Delta=0$, $g=f=0.08\xi$ and $\eta=0.002 \xi$. The lower part is the diagram of the energy levels corresponding to the geometric configuration. The blue line represents the extra driving field, the green line represents the atomic coupling induced by the waveguide, and the yellow dashed line represents the associated dissipation.}
\label{magthree}
\end{centering}
\end{figure*}
In the above section, we find an intriguing phenomena that the giant atom system will generate the atomic entanglement even between the the next-neighbour atom pairs. It is naturally to further explore how to manipulate the degree of entanglement. One of the approach is to induce the phase in the driven field, which serves as the artificial magnetic fields~\cite{IB2012,YY2021,AL2012,AL2016}.
{Due to the cyclic coupling, the phase difference between different atoms can not be eliminated by any gauge transformation. Therefore, we write the driving Hamiltonian in the rotating frame as}
\begin{equation} H_d=\underset{n,m}{\sum}\eta(\sigma_{+}^{(n)}+\Gamma_{+}^{(m)}e^{-i\phi}+{\rm H.c.} ),
\end{equation}
with the strength $\eta$ and the phase $\phi$. Phenomenally, we approximate that the other parts in the effective Hamiltonian and the master equation is not changed by the phase $\phi$.

We begin with the two atom setup, where the modulation of entanglement via the artificial magnetic field (the driving phase cannot be eliminated by canonical transformation) is demonstrated in Fig.~\ref{magtwo} for both giant and small atoms configurations. A general result is that the modulation to the giant atomic configuration is more obvious than that to the small atoms. Therefore, the giant atom system provide us an approach to tune the entanglement with a wider regime via the artificial field.
As we have already discussed in Sec.~\ref{twoatom}, the effective coupling of the two giant atoms is same in cases (II) and (III). However, the presence of the larger individual dissipation of the $B$ atom in case (III) with $J_{2} = 2J$ makes the concurrence decrease rapidly with the modulation of the driving phase, as shown by the purple dashed line in Fig.~\ref{magtwo}.
Futhermore, the two cases listed in Fig.~\ref{smallinteraction} for two small atoms are same with each other, therefore the curves for the small atoms coincide with each other. Case (II) in the giant atom configurations has the similar form as in the small atom configurations, as listed in Fig.~\ref{masterFig} and Fig.~\ref{smallinteraction}. Therefore, the significant difference in the modulation is originated from the waveguide-induced Lamb shift of the giant atoms, which is not achievable in the small atom configurations.

Next, let us move to the system with three atoms and compare the steady state entanglement between the giant and small atoms setups. In the upper panel of Figs.~\ref{magthree} (a)-(d), we illustrate the entanglement for cases (II), (III) in the giant atom setups and cases (I), (II) in the small atom setups, and the corresponding energy level diagrams are also given in the lower panel. In the energy diagram, we define the state $|\psi_{\rm A_{1}}\rangle=|e,g,g\rangle, |\psi_{\rm B_{1}}\rangle=|g,e,g\rangle$, $|\psi_{\rm A_{2}}\rangle=|g,g,e\rangle$ and $|\psi_{0}\rangle=|g,g,g\rangle$. The blue (green) solid arrows represent the driving field (waveguide) induced transition and the yellow dashed arrows are the waveguide induced atomic collective dissipation.

We first discuss the setup with three giant atoms, in which both of the neighbour atomic entanglement [$\mathcal{C}_{ss}(A_1B_1)=\mathcal{C}_{ss}(B_1A_2)$] and the next neighbour atomic entanglement [$\mathcal{C}_{ss}(A_1A_2)$] can be effectively modulated by the artificial magnetic field. This can be explained by the cyclic energy diagram as shown in the lower panel of Figs.~\ref{magthree} (a) and (b).
For case (II) represented by Fig.~\ref{magthree}(a), the presence of the driving field leads to the cyclic diagram to the system. For case (III), the energy diagram in Fig.~\ref{magthree}(b) shows that the waveguide induced transitions form a closed circle, regardless of the driving field. Therefore, the next-neighbour entanglement can be significantly modulated by an artificial magnetic field and even the next-neighbour entanglement can surpass those of neighbour ones, which is exhibited in the shaded regime of the upper panel of Figs.~\ref{magthree} (a-b).

The results for three small atom setups are given in Figs.~\ref{magthree} (c) and (d) for cases (I) and (II), respectively. Here, the cyclic transition is formed with the assistance of the driving field, with the difference being that the state $|\psi_{\rm a_{2}}\rangle$ is excluded from the circle in case (I) [as shown in Fig.~\ref{magthree}(c)] but included in case (II) [as shown in Fig.~\ref{magthree}(d)]. Therefore, the next neighbour entanglement is always zero in the former case but can be manipulated by the artificial magnetic field in the later case. On the other hand, in the small atoms cases, the next-neighbour entanglement is not only smaller than the counterpart in giant atom but also smaller than the neighbour ones.
Therefore, the giant atom system is of great potential to obtain and manipulate the atomic entanglement.

\section{Conclusion}
\label{con}
In this paper, we have proposed an scheme to realize the controllable coupling
and collective dissipation among the giant atoms, which is mediated by the coupled resonator waveguide. In the microwave regime, the working frequency transmission line resonator is in the order of GHz, the photonic hopping strength between the neighbour site $\xi$ can achieve hundreds of MHz~\cite{JK2007,JM2007}.
With the platform of superconducting materials, the controllable coupled resonator waveguide has also been realized by the high-impedance microwave resonators, and by expanding the capacitively coupled lumped-element, the nearest hopping strength has been achieved from 50 MHz to 200 MHz~\cite{PRou2017,MS2022,XZ2023}.
The giant atom were initially realized at the surface acoustic wave platform by coupling a superconducting transmon quantum qubit~\cite{MV2014}. With the development in microwave superconducting materials, the giant atom was also achieved by coupling artificial atoms made of Josephson junctions into superconducting circuits through capacitance or inductance. In such systems, the coupling strength $g$ is much smaller than the hopping strength $\xi$ with the existing technology\cite{ZH2018,BR2010,RM2019}.

In conclusion, we have investigated how to engineer the interaction and entanglement among the driven giant atoms via tuning their geometric configurations.
In the giant atom system, the cyclic coupling drives the system to produce more pronounced next neighbour atomic entanglement. By adjusting the artificial magnetic field, the next neighbour atomic entanglement can be further enhanced and is expected to surpass the neighbour ones in the giant atoms system in principle.
{This entanglement is limited by the fact that the waveguide induced dissipation is in the same order of magnitude as the effective coupling strength. Actually, both of the neighbour and next neighbour atomic entanglement is hopefully enhanced with assistant of some strategies. For example, to induce the direct atomic interaction or design the giant atom array system to support strong entanglement via the quantum phase transition, etc. Since the giant atom setup is mainly realized in superconducting circuits, which is widely used to design kinds of coupling schemes, these proposals are potentially achievable experimentally.}
This next neighbour atomic enhancement can not be found in the small atom setup, and we expect that it is useful in constructing the quantum network and realizing quantum information processing.

\begin{acknowledgments}
Z. Wang is supported by the Science and Technology Development Project of Jilin Province (Grant No. 20230101357JC) and National Science Foundation of China (Grant No. 12375010). X. Wang is supported by the National Natural Science Foundation of China (NSFC; No.~12174303 and Grant No.~11804270), and the Fundamental Research Funds for the Central Universities (No. xzy012023053).

\end{acknowledgments}
\appendix
\addcontentsline{toc}{section}{Appendices}\markboth{APPENDICES}{}
\begin{subappendices}
\section{Master equation for giant-atom setup}
\label{A1}
In this Appendix we outline the derivation of the master equation Eq.~(\ref{masterequation}) which governs the dynamics of the system by considering the coupling resonator waveguide as the environment. In the interaction picture, the interaction Hamiltonian can be rewritten as follow
\begin{eqnarray}
H_{I}&=&g\underset{n}{\sum}(\sigma_{+}^{(n)}E(A_{n},t)e^{i\Omega t}+{\rm h.c.})\nonumber\\
&&+f\underset{m}{\sum}(\Gamma_{+}^{(m)}E(B_{m},t)e^{i\Omega t}+{\rm h.c.}).
\end{eqnarray}
where
\begin{eqnarray}
E(A_{n},t)&=&\frac{1}{\sqrt{N_{c}}}\underset{k,n}{\sum}(a_{k}e^{-i\omega_{k}t}e^{ikx_{n}}+a_{k}e^{-i\omega_{k}t}e^{-ik(x_{n}+t_{A})}),\nonumber\\
E(B_{m},t)&=&\frac{1}{\sqrt{N_{c}}}\underset{k,m}{\sum}(a_{k}e^{-i\omega_{k}t}e^{iky_{m}}+a_{k}e^{-i\omega_{k}t}e^{-ik(y_{m}+t_{B})}).\nonumber
\\
\end{eqnarray}
Under the Markov approximation and in the interaction picture, the formal master equation for the quantum open system reads\cite{HB}
\begin{equation}
\overset{\cdot}{\rho}(t)=-\int_{0}^{\infty}d\tau{\rm Tr_{c}}\{[H_{I}(t),[H_{I}(t-\tau),\rho_{c}\otimes\rho(t)]]\}.
\end{equation}
Since we are working at zero temperature, the waveguide is in its vacuum state initially. Therefore, we will have ${\rm Tr}_c[E^{\dagger}(X,t)E(X,t-\tau)\rho_c]=0$. Back to the Schr$\ddot{o}$dinger picture, the above equation becomes
\begin{eqnarray}
\dot{\rho}&=&\underset{n}{\sum}\underset{m}{\sum}\{-i[\frac{\Omega}{2}\sigma_{z}^{n}+\frac{\Omega}{2}\Gamma_{z}^{m},\rho]\nonumber\\
&&+(A_{11}+A_{11}^{*})\sigma_{-}^{n}\rho\sigma_{+}^{m}-A_{11}\sigma_{+}^{n}\sigma_{-}^{m}\rho-A_{11}^{*}\rho\sigma_{+}^{n}\sigma_{-}^{m}\nonumber\\
&&+(A_{22}+A_{22}^{*})\Gamma_{-}^{n}\rho\Gamma_{+}^{m}-A_{22}\Gamma_{+}^{n}\Gamma_{-}^{m}\rho-A_{22}^{*}\rho\Gamma_{+}^{n}\Gamma_{-}^{m}\nonumber\\
&&+(A_{12}+A_{12}^{*})\sigma_{-}^{n}\rho\Gamma_{+}^{m}-A_{12}\sigma_{+}^{n}\Gamma_{-}^{m}\rho-A_{12}^{*}\rho\sigma_{+}^{n}\Gamma_{-}^{m}\nonumber\\
&&+(A_{21}+A_{21}^{*})\Gamma_{-}^{n}\rho\sigma_{+}^{m}-A_{21}\Gamma_{+}^{n}\sigma_{-}^{m}\rho-A_{21}^{*}\rho\Gamma_{+}^{n}\sigma_{-}^{m}\}.\nonumber
\\
\end{eqnarray}
where
\begin{eqnarray}
A_{11}&=&g^2\int_{0}^{\infty}d\tau e^{i\Omega \tau}{\rm Tr}_{c}(\underset{n}{\sum}\underset{m}{\sum}E(A_n,t)E^{\dagger}(A_m,t-\tau)\rho_{c})\nonumber\\
&=&g^2\underset{n}{\sum}\underset{m}{\sum}\int_{0}^{\infty}d\tau e^{i\Omega \tau}{\rm Tr}_{c}[(E(x_n,t)+E(x_n+t_A,t))\nonumber\\
&&\times(E^{\dagger}(x_m,t-\tau)+E^{\dagger}(x_m+t_A,t-\tau))]\rho_{c})\nonumber\\
&=&g^2\underset{n}{\sum}\underset{m}{\sum}\int_{0}^{\infty}d\tau e^{i\Omega \tau}{\rm Tr}_{c}[E(x_n,t)E^{\dagger}(x_m,t-\tau)\nonumber\\
&&+E(x_n,t)E^{\dagger}(x_m+t_A,t-\tau)\nonumber\\
&&+E(x_n+t_A,t)E^{\dagger}(x_m,t-\tau)\nonumber\\
&&+E(x_n+t_A,t)E^{\dagger}(x_m+t_A,t-\tau)].
\label{A11}
\end{eqnarray}
We can see that the integral in the above equation involves tracing over the four terms. Let's take one of them for example to show our calculation. The first term of the above equation is\cite{GC2016}

\begin{eqnarray}
&&\underset{n}{\sum}\underset{m}{\sum}\int_{0}^{\infty}d\tau e^{i\Omega \tau}{\rm Tr}_{c}E(x_n,t)E^{\dagger}(x_m,t-\tau)\nonumber\\
&=&\underset{n}{\sum}\underset{m}{\sum}\int_{0}^{\infty}d\tau\frac{e^{i\Omega \tau}}{N_{c}}\nonumber\\
&&\times{\rm Tr}[\underset{k,k^{'}}{\sum}e^{-i\omega_{k}t}e^{ik x_n}a_{k}e^{i\omega_{k^{'}}(t-\tau)}e^{-ik^{'} x_m}a_{k^{'}}^{\dagger},\rho_{c}]\nonumber\\
&=&\underset{n}{\sum}\underset{m}{\sum}\int_{0}^{\infty}\frac{d\tau}{N_{c}}\underset{k}{\sum}[e^{-i(\omega_{k}-\Omega)\tau}e^{-ik(x_m-x_n)}]\nonumber\\
&=&\underset{n}{\sum}\underset{m}{\sum}\int_{0}^{\infty}\frac{d\tau}{N_{c}}\sum_{n=0}^{N_c-1}e^{-i\Delta_c}\tau e^{\frac{-2\pi i(x_m-x_n)n_c}{N_c}}e^{2i\xi\cos(\frac{2\pi n_c}{N_c})\tau}\nonumber\\
&=&\underset{n}{\sum}\underset{m}{\sum}\int_{0}^{\infty}d\tau\frac{e^{-i\Delta_c}\tau}{N_{c}}\sum_{n=0}^{N_c-1}e^{\frac{-2\pi i(x_m-x_n)n_c}{N_c}}\nonumber\\
&&\times \sum_{j=-\infty}^{\infty}i^{j}J_{j}(2\xi\tau)e^{i2\pi n_{c}j/N_c}\nonumber\\
&=&\underset{n}{\sum}\underset{m}{\sum}\int_{0}^{\infty}d\tau e^{-i\Delta_c\tau}i^{|x_n-x_m|}J_{|x_n-x_m|}(2\xi\tau)\nonumber\\
&=&\underset{n}{\sum}\underset{m}{\sum}\frac{1}{2\xi}e^{i\frac{\pi}{2}|x_n-x_m|}.
\end{eqnarray}
In the above calculations,we have considered that the giant atom is resonant with the bare cavity $(\Delta_c = \omega_c-\Omega=0)$, and with the help of the following formula
\begin{equation}
\int_0^{\infty}d\tau J_j(a\tau)=\frac{1}{|a|}.
\end{equation}
Therefore, we can complete the calculation of Eq.~(\ref{A11}) and obtain
\begin{eqnarray}
A_{11}^{(n,m)}&=&\frac{1}{2\xi}(2e^{i\frac{\pi}{2}|x_n-x_m|}\nonumber\\
&&+e^{i\frac{\pi}{2}|x_n-x_m-t_A|}+e^{i\frac{\pi}{2}|x_n+t_A-x_m|}).
\end{eqnarray}
Similarly, other coefficients in Eq.(A4) can be obtained.
\begin{eqnarray}
A_{22}^{(n,m)}&=&\frac{1}{2\xi}(2e^{i\frac{\pi}{2}|y_n-y_m|}\nonumber\\
&&+e^{i\frac{\pi}{2}|y_n-y_m-t_B|}+e^{i\frac{\pi}{2}|y_n+t_B-y_m|}).\\
A_{12}^{(n,m)}&=&\frac{1}{2\xi}(e^{i\frac{\pi}{2}|x_n-y_m|}+e^{i\frac{\pi}{2}|x_n-y_m-t_B|}\nonumber\\
&&+e^{i\frac{\pi}{2}|x_n+t_A-y_m|}+e^{i\frac{\pi}{2}|x_n+t_A-y_m-t_B|}).\nonumber\\
\\
A_{21}^{(n,m)}&=&\frac{1}{2\xi}(e^{i\frac{\pi}{2}|y_n-x_m|}+e^{i\frac{\pi}{2}|y_n-y_m-t_A|}\nonumber\\
&&+e^{i\frac{\pi}{2}|y_n+t_B-x_m|}+e^{i\frac{\pi}{2}|y_n+t_B-x_m-t_A|}).\nonumber\\
\
\end{eqnarray}

After regrouping the terms, we will get the final form which is given in Eq.~(\ref{Aij}) of the main text.

\section{Master equation for small-atom setup}
\label{A2}

\begin{figure}
\centering
\includegraphics[width=1\columnwidth]{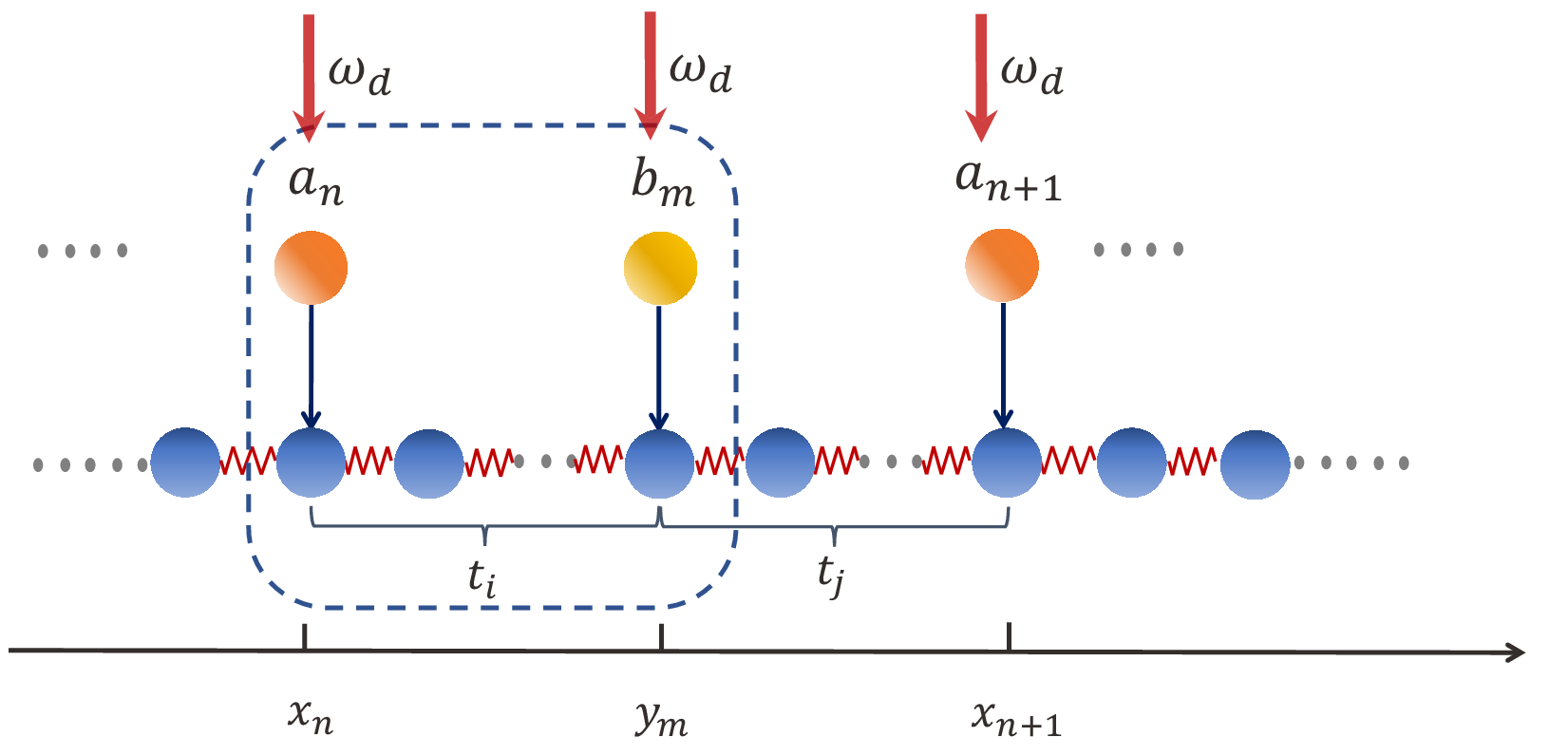}
\caption{Small atoms coupled to a 1D coupled-resonator waveguide. We label the odd number of small atoms as a (orange ball) and the even number of small atoms as $b$ (yellow ball).}
\label{smalldevice}
\end{figure}

In the main text, we compare and discuss the configuration of giant atoms with that of small atoms. Here, we provide a specific model and master equation form for the considered small atom configuration, and briefly discuss them. Similarly, we consider the interaction between a small atomic array composed of two-level atoms and  a one-dimensional coupled-resonator waveguide.
We consider the array composed of small atoms, where two neighboring small atoms form a protocell, labeled $a$ and $b$, respectively, with different the intra-cell atomic distance and the extra-cell atomic distance that can be adjusted, sketched in Fig.~\ref{smalldevice}. Giant atoms can be coupled to the waveguide through multiple points, while each small atom can only be coupled to the waveguide via one point. In the small atom configuration, the Hamiltonian of the system is written as
\begin{eqnarray}
H_{c}
&=&\omega_{c}\underset{j}{\sum}a_{j}^{\dagger}a_{j}-\xi\underset{j}{\sum}(a_{j+1}^{\dagger}a_{j}+a_{j}^{\dagger}a_{j+1}),\\
H
&=&H_{c}+\frac{\Omega}{2}\underset{n}{\sum}\nu_{z}^{(n)}+\frac{\Omega}{2}\underset{m}{\sum}\tau_{z}^{(m)}\nonumber \\
&&+\underset{n,m}{\sum}[g a_{{x_{n}}}^{\dagger}\nu_{-}^{(n)}+f a_{{y_{m}}}^{\dagger}\tau_{-}^{(m)}+{\rm H.c.}].
\
\end{eqnarray}
Here, $x_{n}$ and $y_{m}$ are the coupling site between the small atoms and the one-dimensional coupled-resonator waveguide. The parameter $t_{i}$, $t_{j}$ are the intra-cell atomic distance and the extra-cell atomic distance, respectively. After the same derivation process, we can obtain the master equation for small atoms.
\begin{eqnarray}
\overset{\cdot}{\rho}&=&-i[\mathcal{H},\rho]\nonumber\\
&&+\underset{n,m}{\sum}[g^{2}U_{11}^{(n,m)}(2\nu_{-}^{(n)}\rho\nu_{+}^{(m)}-\nu_{+}^{(m)}\nu_{-}^{(n)}\rho-\rho\nu_{+}^{(n)}\nu_{-}^{(m)})\nonumber\\
&&+f^{2}U_{22}^{(n,m)}(2\tau_{-}^{(n)}\rho\tau_{+}^{(m)}-\tau_{+}^{(n)}\tau_{-}^{(m)}\rho-\rho\tau_{+}^{(n)}\tau_{-}^{(m)})\nonumber\\
&&+gfU_{12}^{(n,m)}(2\nu_{-}^{(n)}\rho\tau_{+}^{(m)}-\nu_{+}^{(n)}\tau_{-}^{(m)}\rho-\rho\nu_{+}^{(n)}\tau_{-}^{(m)})\nonumber\\
&&+gfU_{21}^{(n,m)}(2\tau_{-}^{(n)}\rho\nu_{+}^{(m)}-\tau_{+}^{(n)}\nu_{-}^{(m)}\rho-\rho\tau_{+}^{(n)}\nu_{-}^{(m)})].\nonumber\\
\
\end{eqnarray}
where the coherent coupling between the atoms is
\begin{eqnarray}
\mathcal{H}&=&\underset{n,m}{\sum}(\frac{\Omega}{2}\nu_{z}^{(n)}+\frac{\Omega}{2}\tau_{z}^{(m)})\nonumber\\ &&+\underset{n,m}{\sum}[g^{2}I_{11}\nu_{+}^{(n)}\nu_{-}^{(m)}+f^{2}I_{22}\tau_{+}^{(n)}\tau_{-}^{(m)}]\nonumber\\ &&+\underset{n,m}{\sum}[gf(I_{12}\nu_{+}^{(n)}\tau_{-}^{(m)}+I_{21}\tau_{+}^{(n)}\nu_{-}^{(m)})].
\end{eqnarray}
In the above equations, we have defined $U_{ij}=\mathrm{Re}(A_{ij}),I_{ij}=\mathrm{Im}(A_{ij})(i,j=1,2)$, where
\begin{eqnarray}
A_{11}^{(n,m)}&=&\frac{1}{2\xi}e^{i\frac{\pi}{2}|x_n-x_m|},\nonumber\\
A_{11}^{(n,m)}&=&\frac{1}{2\xi}e^{i\frac{\pi}{2}|y_n-y_m|},\nonumber\\
A_{12}^{(n,m)}&=&A_{21}^{(n,m)}=\frac{1}{2\xi}e^{i\frac{\pi}{2}|x_{n}-y_{m}|}.
\end{eqnarray}

\begin{figure}
\centering
\includegraphics[width=1\columnwidth]{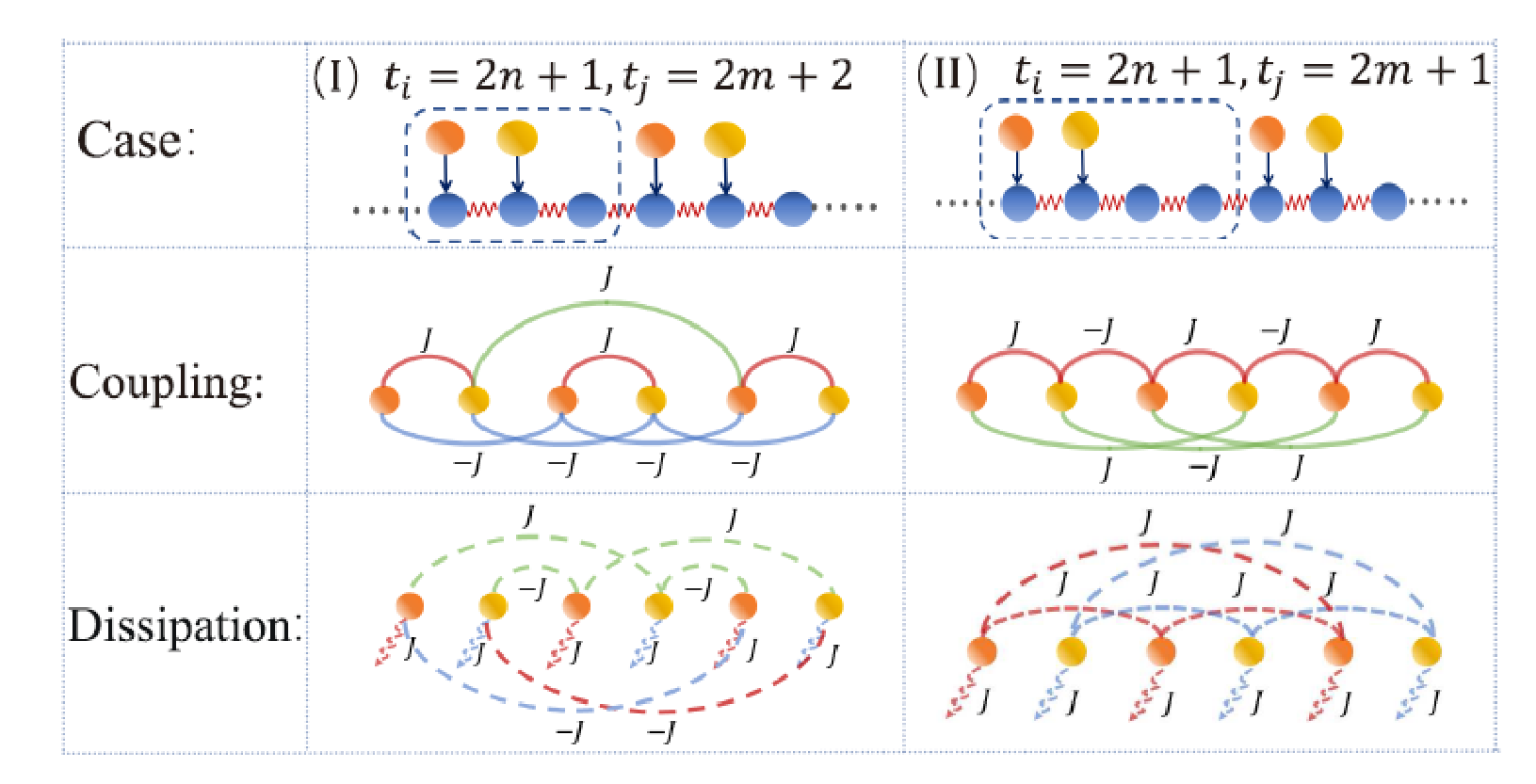}
\caption{The effective couplings and dissipations for two different cases. The first line is the simple illustration of different systems. The second line is a schematic diagram of the corresponding effective coherent interaction. The last line shows the diagram of effective dissipation.}
\label{smallinteraction}
\end{figure}

The form of interaction between small atoms is also related to the spacing between small atoms. For convenience, we fix the intracellular distance as $t_{i}=1$. By changing the intercellular distance $t_{j}$, we find that there are two cases of interactions for small atoms, as shown in the Fig.~\ref{smallinteraction}. In Fig.~\ref{smallinteraction}, we present the geometric configurations of two cases of interaction and their corresponding associated dissipation. For the two small atom configurations, it can be visually seen from the Fig.~\ref{smallinteraction} that there is no cyclic coupling structure as the giant atom configuration.

With the two atoms setup, the master equation for the (I) and (II) cases becomes
\begin{eqnarray}
\dot{\rho}&=&-i[\mathcal{H}_{\rm I}+H_{d},\rho]
+JD_{[\nu_{+},\nu_{-}]\rho}+JD_{[\tau_{+},\Gamma_{-}]\rho},\\
\mathcal{H}_{\rm {I}}&=&\Delta\nu_{z}+\Delta\Gamma_{z}
+J(\nu_{+}\tau_{-}+\tau_{+}\nu_{-}).
\end{eqnarray}
and
\begin{eqnarray}
\dot{\rho}&=&-i[\mathcal{H}_{\rm II}+H_{d},\rho]
+JD_{[\nu_{+},\nu_{-}]\rho}+JD_{[\tau_{+},\tau_{-}]\rho},\nonumber\\\\
\mathcal{H}_{\rm {II}}&=&\Delta\nu_{z}+\Delta\tau_{z}+J(\nu_{+}\tau_{-}+\tau_{+}\nu_{-}).
\end{eqnarray}
where $\mathcal{H}$ is the effective Hamiltonian, and $D_{[O_1,O_2]\rho}=2O_2\rho O_1 -\rho O_1O_2-O_1O_2 \rho$. In the rotating frame, the driving Hamiltonian is written as $H_d=\eta(\nu_{+}+\tau_{+}+{\rm H.c.})$.

The master equation for the three atoms setup are $\dot{\rho}=-i[\mathcal{H}_i,\rho]+D_i\rho\,(i={\rm I}, {\rm II})$, where the corresponding effective Hamiltonians and the dissipators in the rotating frame are respectively
\begin{eqnarray}
\mathcal{H}_{\rm I}&=&\Delta(\nu_{z}^{(1)}+\nu_{z}^{(2)})
+\Delta\tau_{z}^{(1)}\nonumber\\
&&+(J\nu_{+}^{(1)}\tau_{-}^{(1)}-J\nu_{+}^{(1)}\nu_{-}^{(2)}+{\rm H.c.}),\\
\label{SmallHThreeI}
D_{\rm I}\rho&=&JD_{[\nu_{+}^{(1)},\nu_{-}^{(1)}]\rho}+JD_{[\nu_{+}^{(2)},\nu_{-}^{(2)}]\rho}+JD_{[\tau_{+}^{(1)},\tau_{-}^{(1)}]\rho}\nonumber\\
&&-J\left(D_{[\tau_{+}^{(1)},\nu_{-}^{(2)}]\rho}+D_{[\nu_{+}^{(2)},\tau_{-}^{(1)}]\rho}\right).
\end{eqnarray}
and
\begin{eqnarray}
\mathcal{H}_{\rm II}&=&\Delta(\nu_{z}^{(1)}+\nu_{z}^{(2)})
+\Delta\tau_{z}^{(1)}\nonumber\\
&&+(J\nu_{+}^{(1)}\tau_{-}^{(1)}-J\tau_{+}^{(1)}\nu_{-}^{(2)}+{\rm H.c.}),\\
\label{SmallHThreeII}
D_{\rm II}\rho&=&JD_{[\nu_{+}^{(1)},\nu_{-}^{(1)}]\rho}+JD_{[\nu_{+}^{(2)},\nu_{-}^{(2)}]\rho}+JD_{[\tau_{+}^{(1)},\tau_{-}^{(1)}]\rho}\nonumber\\
&&-J\left(D_{[\nu_{+}^{(1)},\nu_{-}^{(2)}]\rho}+D_{[\nu_{+}^{(2)},\nu_{-}^{(1)}]\rho}\right).
\end{eqnarray}

\end{subappendices}


\begin{thebibliography}{99}

\bibitem{PF2019}P. Forn-D\'{\i}az, L. Lamata, E. Rico, J. Kono, and E. Solano, Rev. Mod. Phys. {\bf 91}, 025005 (2019).

\bibitem{RG2021}R. Gutzler, M. Garg, C. R. Ast, K. Kuhnke, and K. Kern, Nat. Rev. Phys. {\bf 3}, 441 (2021).
%
\bibitem{DF}D. F. Walls and G. J. Milburn, \emph{Quantum Optics}, 2nd ed. (Spring, Berling, 2008).
%
%
\bibitem{MV2014}M. V. Gustafsson, T. Aref, A. F. Kockum, M. K. Ekstr{\"o}m, G. Johansson, and P. Delsing, Science {\bf 346}, 207 (2014).
%
%
\bibitem{AF2014}A. Frisk Kockum, P. Delsing, and G. Johansson, Phys. Rev. A {\bf 90}, 013837 (2014).
%
\bibitem{RM2017}R. Manenti, A. F. Kockum, A. Patterson, T. Behrle, J. Rahamim, G. Tancredi, F. Nori, and P. J. Leek, Nat. Commun. {\bf 8}, 975 (2017).
%
\bibitem{AN2017}A. Noguchi, R. Yamazaki, Y. Tabuchi, and Y. Nakamura, Phys. Rev. Lett. {\bf 119}, 180505 (2017).
%
\bibitem{KJ2018}K. J. Satzinger, Y. P. Zhong, H.-S. Chang, G. A. Peairs, A. Bienfait, M. H. Chou, A. Y. Cleland, C. R. Conner, \'{E}. Dumur, J. Grebel, I. Gutierrez, B. H. November, R. G. Povey, S. J. Whiteley, D. D. Awschalom, D. I. Schuster, and A. N. Cleland, Nature (London) {\bf 563}, 661 (2018).
%
\bibitem{AA2019}A. Ask, M. Ekstr{\"o}m, P. Delsing, and G. Johansson, Phys. Rev. A {\bf 99}, 013840 (2019).
%
\bibitem{AF2021}A. F. Kockum, Quantum optics with giant atoms the first five years, P125, in Mathematics for Industry (Springer Singapore, 2021).
%
\bibitem{BK2020}B. Kannan, M. Ruckriegel, D. Campbell, A. F. Kockum, J. Braum{\"u}ller, D. Kim, M. Kjaergaard, P. Krantz, A. Melville, B. M. Niedzielski, A. Veps{\"a}l{\"a}inen, R. Winik, J. Yoder, F. Nori, T. P. Orlando, S. Gustavsson, and W. D. Oliver, Nature (London) {\bf 583}, 775 (2020).
%
\bibitem{YT2022}Y. T. Chen, L. Du, L. Guo, Z. Wang, Y. Zhang, Y. Li, and J. H. Wu, Commun. Phys. {\bf 5}, 215 (2022).
%
\bibitem{XL2022}X.-L. Yin, Y.-H. Liu, J.-F. Huang, and J.-Q. Liao,  Phys. Rev. A {\bf 106}, 013715 (2022).
%
%
\bibitem{LG2017}L. Guo, A. L. Grimsmo, A. F. Kockum, M. Pletyukhov, and G. Johansson, Phys. Rev. A {\bf 95}, 053821 (2017).
%
\bibitem{LD2021}L. Du, M. Cai, J. Wu, Z. Wang, and Y. Li, Phys. Rev. A {\bf 103}, 053701 (2021).
%
\bibitem{QY2023}Q.-Y. Qiu, Y. Wu, and X.-Y. Lu, Sci. China Phys. Mech. Astron. {\bf 66}, 224212 (2023).
%
%
\bibitem{LG2020}L. Guo, A. F. Kockum, F. Marquardt, and G. Johansson, Phys. Rev. Research {\bf 2}, 043014 (2020).
%
\bibitem{WZ2020}W. Zhao and Z. Wang, Phys. Rev. A {\bf 101}, 053855 (2020).
%
\bibitem{XW2021}X. Wang, T. Liu, A. F. Kockum, H.-R. Li, and F. Nori, Phys. Rev. Lett. {\bf 126}, 043602 (2021).
%
\bibitem{WC2022}W. Cheng, Z. Wang, and Y.-x. Liu, Phys. Rev. A {\bf 106}, 033522 (2022).
%
\bibitem{KH2023}K. H. Lim, W.-K. Mok, and L.-C. Kwek, Phys. Rev. A {\bf 107}, 023716 (2023).
%
%
%
\bibitem{GA2019}G. Andersson, B. Suri, L. Guo, T. Aref, and P. Delsing, Nat. Phys. {\bf 15}, 1123 (2019).
%
\bibitem{SG2020}S. Guo, Y. Wang, T. Purdy, and J. Taylor, Phys. Rev. A {\bf 102}, 033706 (2020).
%
\bibitem{XY2022}X. Yin, W. Luo, and J. Liao, Phys. Rev. A {\bf 106}, 063703 (2022).
%
\bibitem{LD2023}L. Du, L. Guo, and Y. Li, Phys. Rev. A {\bf 107}, 023705 (2023).
%
\bibitem{XJ2023}X. Zhang, C. Liu, Z. Gong, and Z. Wang, Phys. Rev. A {\bf 108}, 013704 (2023).
%
%
\bibitem{AF2018}A. F. Kockum, G. Johansson, and F. Nori, Phys. Rev. Lett. {\bf 120}, 140404 (2018).
%
\bibitem{AC2020}A. Carollo, D. Cilluffo, and F. Ciccarello, Phys. Rev. Research {\bf 2}, 043184 (2020).
%
%
\bibitem{XW2022}X. Wang and H.-R. Li, Quantum Sci. Technol. {\bf 7}, 035007 (2022).
%
\bibitem{CJ2023}C. Joshi, F. Yang, and M. Mirhosseini, Phys. Rev. X {\bf 13}, 021039 (2023).
%
{\bibitem{XWarxiv}X. Wang, J.-Q Li, Z. Wang, A. F. Kockum, L. Du, T. Liu, and F. Nori, arxiv: 2404.09829 (2024).}
%
%
\bibitem{PY2019}P. Y. Wen, K. Lin, A. F. Kockum, B. Suri, H. Ian, J. C. Chen, S. Y. Mao, C. C. Chiu, P. Delsing, F. Nori, G. Liu, and I. Hoi, Phys. Rev. Lett. {\bf 123}, 233602 (2019).
%
\bibitem{AM2021}A. M. Vadiraj, Andreas Ask, T. G. McConkey, I. Nsanzineza, C. W. Sandbo Chang, A. F. Kockum, and C. M. Wilson, Phys. Rev. A {\bf 103}, 023710 (2021).
%
%
%
\bibitem{SL2020}S. Longhi, Opt. Lett. {\bf 45}, 3017 (2020).
%
%
\bibitem{ZQ2022}Z. Q. Wang, Y. P. Wang, J. Yao, R. C. Shen, W. J. Wu, J. Qian, J. Li, S. Y. Zhu, and J. Q. You, Nat. Commun. {\bf 13}, 7580 (2022).
%
%
\bibitem{AG2019}A. Gonz\'{a}lez-Tudela, C. S\'{a}nchez Munoz, and J. I. Cirac, Phys. Rev. Lett. {\bf 122}, 203603 (2019).
%
%
%
\bibitem{LD2022}L. Du, Y. Zhang, J.-H. Wu, A. F. Kockum, and Y. Li, Phys. Rev. Lett. {\bf 128}, 223602 (2022).
%
\bibitem{HX2022}H. Xiao, L. Wang, Z.-H. Li, X. Chen, and L. Yuan, npj Quantum Infom. {\bf 8}, 80 (2022).
%
%
\bibitem{XW2023arxiv}X. Wang, H.-B. Zhu, T. Liu, and F. Nori, arxiv:2304.10710.
%
%
\bibitem{AS2023}A. Soro, C. S. Mu\~{n}oz, and A. F. Kockum, Phys. Rev. A {\bf 107}, 013710 (2023).
%
%
\bibitem{SL2021}S. L. Feng and W. Z. Jia, Phys. Rev. A {\bf 104}, 063712 (2021).
%
\bibitem{AS2022}A. Soro, and A. F. Kockum, Phys. Rev. A {\bf 105}, 023712 (2022).
%
{\bibitem{ERarxiv}E. R. Ingelsten, A. F. Kockum, and A. Soro,
arxiv: 2402.10879 (2024).}

{\bibitem{LLarxiv}L. Leonforte, X. Sun, D. Valenti, B. Spagnolo, F. Illuminati, A. Carollo, and F. Ciccarello, arxiv: 2402.10275 (2024).}
%
\bibitem{DC2020}D. Cilluffo, A. Carollo, S. Lorenzo, J. A. Gross, G. M. Palma, and F. Ciccarello, Phys. Rev. Research {\bf }2, 043070 (2020).
%
\bibitem{QY2021}Q. Y. Cai and W. Z. Jia, Phys. Rev. A {\bf 104}, 033710 (2021).
%
\bibitem{HY2021}H. Yu, Z. Wang, and J.-H. Wu, Phys. Rev. A {\bf 104}, 013720 (2021).
%
\bibitem{DD2022}D. D. Noachtar, J. Korzer, R. H. Jonsson, Phys. Rev. A {\bf 106}, 013702 (2022).
%
\bibitem{AC2023}A. C. Santos and R. Bachelard, Phys. Rev. Lett. {\bf 130}, 053601 (2023).
%
%
\bibitem{LM2013}L. M. Sieberer, S. D. Huber, E. Altman, and S. Diehl, Phys. Rev. Lett. {\bf 110}, 195301 (2013).
%
\bibitem{TL2017}T. Leleu, Y. Yamamoto, S. Utsunomiya, and K. Aihara, Phys. Rev. E {\bf 95}, 022118 (2017).
%
\bibitem{TF2018}T. Fink, A. Schade, S. H{\"o}fling, C. Scheider, and A. Imamoglu, Nat. Phys. {\bf 14}, 365 (2018).
%
\bibitem{FM2018}F. Minganti, A. Biella, N. Bartolo, and C. Ciuti, Phys. Rev. A {\bf 98}, 042118 (2018).
%
\bibitem{LR2022}L. R. Bakker, M. S. Bahovadinov, D. V. Kurlov, V. Gritsev, A. K. Fedorov, and Dmitry O. Krimer, Phys. Rev. Lett. {\bf 129}, 250401 (2022).
%
\bibitem{NP2022}N. Pernet, P. St-Jean, D. D. Solnyshkov, G. Malpuech, N. C. Zambon, Q. Fontaine, B. Real, O. Jamadi, A. Lemaitre, M. Morassi, L. L. Gratiet, T. Baptiste, A. Harouri, I. Sagnes, A. Amo, S. Ravets, and J. Bloch, Nat. Phys. {\bf 18}, 678 (2022).
%
%
\bibitem{XY2023}X. Yin and J.Liao, Phys. Rev. A {\bf 108}, 023728 (2023).
%
%
%
\bibitem{MA2013}M. Aidelsburger, M. Atala, M. Lohse, J. T. Barreiro, B. Paredes, and I. Bloch. Phys. Rev. Lett. {\bf 111}, 185301 (2013).
%
\bibitem{LT2014}L. Tzuang, K. Fang, P. Nussenzveig, S. Fan, and M. Lipson, Nat. Photon. {\bf 8}, 701 (2014).
%
\bibitem{PR2017}P. Roushan, C. Neill, A. Megrant, Y. Chen, R. Babbush, R. Barends, B. Campbell, Z. Chen, B. Chiaro, A. Dunsworth \emph{et al.}, Nat. Phys. {\bf 13}, 146 (2017).
%
%
\bibitem{IB2012}I. Bloch, J. Dalibard, and S. Nascimbene, Nat. Phys. {\bf 8}, 267 (2012).
%
%
\bibitem{AL2012} A. L. C. Hayward, A. M. Martin, and A. D. Greentree, Phys. Rev. Lett. {\bf 108}, 223602 (2012).
%
\bibitem{AL2016}A. L. C. Hayward and A. M. Martin, Phys. Rev. A {\bf 93}, 023828 (2016).
%
\bibitem{YY2021}Y.-Y. Zhang, Z.-X. Hu, L. Fu, H.-G. Luo, H. Pu, and X.-F. Zhang, Phys. Rev. Lett. {\bf 127}, 063602 (2021).
%
%
\bibitem{GC2016}G. Calaj\'{o}, F. Ciccarello, D. Chang, and P. Rabl, Phys. Rev. A {\bf 93}, 033833 (2016).
%
\bibitem{HB} H. Breuer and F. Petruccione, The Theory of Open Quantum Systems (Oxford University Press, Oxford, 2002).
%
\bibitem{SA1997}S. A. Hill and W. K. Wootters, Phys. Rev. Lett. {\bf 78}, 5022 (1997).
%
%
\bibitem{JK2007}J. Koch, T. M. Yu, J. Gambetta, A. A. Houck, D. I. Schuster,
J. Majer, A. Blais, M. H. Devoret, S. M. Girvin, and R. J.
Schoelkopf, Phys. Rev. A {\bf 76}, 042319 (2007).
%
\bibitem{JM2007}J. Majer, J. M. Chow, J. M. Gambetta, J. Koch, B. R. Johnson,
J. A. Schreier, L. Frunzio, D. I. Schuster, A. A. Houck, A.
Wallraff, A. Blais, M. H. Devoret, S. M. Girvin, and R. J.
Schoelkopf, Nature (London) {\bf 449}, 443 (2007).
%
%
\bibitem{PRou2017}P. Roushan, C. Neill, J. Tangpanitanon, V. M. Bastidas, A. Megrant, R. Barends, Y. Chen, Z. Chen, B. Chiaro, A. Dunsworth, A. Fowler, B. Foxen, M. Giustina, E. Jeffrey, J. Kelly, E. Lucero, J. Mutus, M. Neeley, C. Quintana, D. Sank, A. Vainsencher, J. Wenner, T. White, H. Neven, D. G. Angelakis, and J. Martinis, Science {\bf 358}, 1175 (2017).
%
\bibitem{MS2022}M. Scigliuzzo, G. Calaj\'{o}, F. Ciccarello, D. P. Lozano, A. Bengtsson, P. Scarlino, A. Wallraff, D. Chang, P. Delsing, and S. Gasparinetti, Phys. Rev. X {\bf 12}, 031036 (2022).
%
\bibitem{XZ2023}X. Zhang, E. Kim, D. K. Mark, S. Choi, and O. Painter, Science {\bf 379}, 278 (2023).
%
%
\bibitem{BR2010}B. R. Johnson, M. D. Reed, A. A. Houck, D. I. Schuster, L. S.
Bishop, E. Ginossar, J. M. Gambetta, L. DiCarlo, L. Frunzio,
S. M. Girvin, and R. J. Schoelkopf, Nat. Phys.
{\bf 6}, 663 (2010).
%
\bibitem{ZH2018}Z. H. Peng, J. H. Ding, Y. Zhou, L. L. Ying, Z. Wang, L. Zhou,
L. M. Kuang, Y.-x. Liu, O. V. Astafiev, and J. S. Tsai, Phys. Rev. A {\bf 97}, 063809 (2018).
%
%
\bibitem{RM2019}R. Ma, B. Saxberg, C. Owens, N. Leung, Y. Lu, J. Simon,
and D. I. Schuster,  Nature (London) {\bf 566}, 51 (2019).
%
%
%
\end{thebibliography}
\end{document}